\documentclass[twocolumn,epjc3]{svjour3mod}
\usepackage{hyperref}
\usepackage{graphicx,colordvi}
\usepackage{amsmath,amssymb,slashed,cite}
\usepackage{booktabs,tabulary}
\usepackage{dsfont}
\usepackage{color}
\usepackage{epstopdf}  

\newcommand{\be}{\begin{equation}}
\newcommand{\ee}{\end{equation}}
\newcommand{\no}{\notag \\}
\newcommand{\fs}{\,.}
\newcommand{\co}{\,,}

\newcommand{\indR}{R}
\newcommand{\indL}{L}
\newcommand{\indT}{T}
\newcommand{\indP}{P}
\newcommand{\qbar}{\bar q}
\newcommand{\ubar}{\bar u}
\newcommand{\dbar}{\bar d}

\newcommand{\MeV}{\,\text{MeV}}
\newcommand{\GeV}{\,\text{GeV}}
\renewcommand{\Im}{\text{Im}\,}
\newcommand{\inel}{\text{inel}}
\newcommand{\el}{\text{el}}
\newcommand{\disp}{\text{disp}}
\newcommand{\ct}{\text{ct}}
\newcommand{\ret}{\text{ret}}
\newcommand{\adv}{\text{adv}}
\newcommand{\fig}{figure}
\renewcommand{\sec}{section}
\newcommand{\subsec}{subsection}

\newcommand{\app}{appendix}
\newcommand{\nuth}{\nu_\text{th}}
\newcommand{\alphaem}{\alpha_\text{em}}
\newcommand{\mN}{m}
\newcommand{\mpi}{M_\pi}
\newcommand{\mg}{m}

\newcommand{\bsp}{\begin{sloppypar}}
\newcommand{\esp}{\end{sloppypar}}
\newcommand{\toright}[1]{\hspace*{\fill}{\footnotesize{#1}}}
 
\RequirePackage{graphicx}

\journalname{Eur. Phys. J. C}
\begin{document}

\title{\toright{\textnormal{INT-PUB-15-027}}\\Cottingham formula and nucleon polarizabilities}
\author{J.\ Gasser\thanksref{addr1}
	\and
	M.\ Hoferichter\thanksref{addr1,addr2,addr3,addr4}
        \and
        H.\ Leutwyler\thanksref{addr1}
        \and
        A.\ Rusetsky\thanksref{addr5} 
}
\institute{Albert Einstein Center for Fundamental Physics, Institut f\"ur theoretische Physik, Universit\"at Bern, Sidlerstrasse 5, CH--3012 Bern, Switzerland\label{addr1}
\and
Institut f\"ur Kernphysik, Technische Universit\"at Darmstadt, D--64289 Darmstadt, Germany\label{addr2}
\and
ExtreMe Matter Institute EMMI, GSI Helmholtzzentrum f\"ur Schwerionenforschung GmbH, 
D--64291 Darmstadt, Germany\label{addr3}
\and
Institute for Nuclear Theory, University of Washington, Seattle, WA 98195-1550, USA\label{addr4}
\and
Helmholtz-Institut f\"ur Strahlen- und Kernphysik (Theorie) and Bethe Center for
 Theoretical Physics, Universit\"at Bonn, Nussallee 14-16, D--53115 Bonn, Germany \label{addr5}
}

\date{}

\maketitle

\begin{abstract}
\bsp
The difference between the electromagnetic self-energies of proton and neutron can be calculated with the Cottingham formula, which expresses the self-energies as an integral over the electroproduction cross sections -- provided the nucleon matrix elements of the current
commutator do not contain a fixed pole. We show that, under the same proviso, the subtraction function occurring in the dispersive representation of the virtual Compton forward scattering amplitude is determined by the cross sections. The representation in particular leads to a parameter-free sum rule for the nucleon polarizabilities. We evaluate the sum rule for the difference between the electric polarizabilities of proton and neutron by means of the available parameterizations of the data and compare the result with experiment.
\esp

\end{abstract}

\tableofcontents

\section{Introduction}
\label{sec:Introduction}

\bsp
The mass difference between proton and neutron had been puzzling for a long time. Ever since Heisenberg had introduced isospin symmetry to explain the near degeneracy of  these two levels~\cite{Heisenberg}, it was taken for granted that the strong interaction is invariant under isospin rotations and that the mass difference is of electromagnetic origin. In this framework, it was difficult, however, to understand the experimental fact that the neutral particle is heavier than the charged one. A first step towards a resolution of the paradox was taken by Coleman and Glashow, who introduced the tadpole dominance hypothesis~\cite{Coleman:1961jn,Coleman:1963pj}, which associates the bulk of the electromagnetic self-energies with an octet operator. The origin of the tadpole remained mysterious, however. The puzzle was solved only in 1975, when it was realized that the strong interaction does not conserve isospin, because the masses of the up- and down-quarks strongly differ~\cite{GL1975}. The crude estimates for the ratios of the three lightest quark masses obtained in that work, $m_u/m_d\simeq 0.67$, $m_s/m_d\simeq 22.5$, have in the meantime been improved considerably. In particular, Weinberg~\cite{Weinberg1977} pointed out that in the chiral limit, the Dashen theorem provides an independent estimate of the quark mass ratios, as it determines the electromagnetic self-energies of the kaons in terms of those of the pions. Neglecting higher orders in the expansion in powers of $m_u,m_d$, and $m_s$, he obtained the estimate $m_u/m_d\simeq 0.56$, $m_s/m_d\simeq 20.1$. Also, the decay $\eta\rightarrow 3\pi$ turned out to be a very sensitive probe of isospin breaking~\cite{Kambor:1995yc,Anisovich:1996tx,Bijnens:2007pr,Schneider:2010hs,Kampf:2011wr}. The quark mass ratios obtained from that source also confirmed the picture.  According to the most recent edition of the FLAG review~\cite{FLAG2014}, the current lattice averages are $m_u/m_d=0.46(3)$, $m_s/m_d= 20.0(5)$.
\esp

\subsection{Cottingham formula, dispersion relations}
\label{subsec:Dispersion theory}

\bsp
The analysis of~\cite{GL1975} relies on the Cottingham formula~\cite{Cottingham}, which invokes dispersion relations to relate the spin-averaged nucleon matrix elements of the time-ordered product, $\langle p|Tj^\mu(x)j^\nu(y)|p\rangle$, to those of the commutator of the electromagnetic current, $\langle p|[j^\mu(x),j^\nu(y)]|p\rangle$. Lorentz invariance and current conservation determine the Fourier transforms of these matrix elements in terms of two invariant amplitudes, which only depend on the two variables $\nu=p\cdot q/\mN$ and $q^2$, where $\mN$ is the nucleon mass and $q$ the photon momentum. We stick to the notation used in~\cite{GL1975} and denote the invariant amplitudes by  $T_1(\nu,q^2), T_2(\nu,q^2)$ and
$V_1(\nu,q^2), V_2(\nu,q^2)$, respectively. Explicit formulae that specify the matrix elements $\langle p|Tj^\mu(x)j^\nu(y)|p\rangle$ and $\langle p|[j^\mu(x),j^\nu(y)]|p\rangle$ in terms of the invariant amplitudes are listed in \app~\ref{app:Notation}, where we also exhibit the relations between the structure functions $V_1(\nu,q^2)$, $V_2(\nu,q^2)$ and the cross sections $\sigma_{\indT}$ and $\sigma_{\indL}$ of electron scattering.
\esp

\bsp
In the space-like region and for $\nu \geq 0$, the structure functions represent the imaginary parts of the time-ordered amplitudes:
\begin{align}
\label{eq:Im}
\Im T_1(\nu,q^2)&=\pi V_1(\nu,q^2)\co\no
\Im T_2(\nu,q^2)&=\pi V_2(\nu,q^2)\co\quad \nu\geq 0\co\, q^2\leq 0\fs
\end{align}
While the functions $V_1(\nu,q^2), V_2(\nu,q^2)$ are odd under $\nu\rightarrow -\nu$, the time-ordered amplitudes $T_1(\nu,q^2), T_2(\nu,q^2)$ are even. In view of the contributions arising from Regge exchange,  $V_1(\nu,q^2) \sim \nu^\alpha$, $V_2(\nu,q^2)\sim \nu^{\alpha-2}$, only $T_2$ obeys an unsubtracted dispersion relation, while for $T_1$ a subtraction is needed.\footnote{It was even suggested that the subtraction term might solve the notorious puzzle with the proton-neutron mass difference, as it could dominate over the remainder and explain why the neutron is heavier than the proton~\cite{Harari}. In hindsight, we know better.} For $q^2<0$, the dispersion relations thus take the form
\begin{align} 
T_1(\nu,q^2)&= S_1(q^2)+2\nu^2\int_0^\infty \frac{d\nu'}{\nu'} \,\frac{V_1(\nu',q^2)}{\nu'^2-\nu^2-i\epsilon}\co\no
 T_2(\nu,q^2)&=  2\int_0^\infty d\nu'\, \nu' \,\frac{ V_2(\nu',q^2)}{\nu'^2-\nu^2-i\epsilon}\fs
 \label{eq:T1}
 \end{align}
 The formulae hold in the cut $\nu$-plane; the upper and lower half-planes are glued together along the interval $|\nu|<Q^2/2\mN$ of the real axis (throughout, we use $Q^2\equiv -q^2$ whenever this is convenient). As illustrated with the discussion in \app~\ref{app:Comments}, it is important that kinematic singularities, zeros, and constraints be avoided -- throughout this paper, we work with the amplitudes defined in \app~\ref{app:Notation}, which are free of these~\cite{Bardeen,Tarrach,Drechsel}. 
 \esp

We refer to $S_1(q^2)$ as the subtraction function. It represents the value of the amplitude $T_1(\nu,q^2)$ at $\nu=0$. For later use we introduce the analogous notation also for $T_2(\nu,q^2)$:
\be 
S_1(q^2)\equiv T_1(0,q^2)\co\quad S_2(q^2)\equiv T_2(0,q^2)\fs
\ee

\subsection{Reggeons and fixed poles}
\label{subsec:Reggeons}

\bsp
In~\cite{GL1975} it is assumed that the asymptotic behaviour is determined by Reggeon exchange. The contribution of a Regge pole to a scattering amplitude at large center-of-mass energy squared $s$ and small momentum transfer $t\leq 0$ has the form (see e.g.~\cite{CCL}):
\be 
T(s,t) = - \frac{\pi \beta_{\alpha}(t)}{\sin\pi \alpha(t) } 
\{\exp[-i\pi\alpha(t)]+\tau\}s^{\alpha(t)}\co
\label{eq:T1RStandard}
\ee
where $\alpha(t)$ and $\beta(t)$ denote the trajectory and the residue,
respectively, and $\tau$ is the signature. In the context of the present paper, we are concerned with $t=0$ and $\tau=1$. The continuation of the asymptotic formula~\eqref{eq:T1RStandard} to low energies  is not unique. For definiteness, we work with the representation  
\begin{align}
T^R_1(\nu,q^2) &= -\sum_{\alpha>0} \frac{\pi \beta_{\alpha}(Q^2)}{\sin\pi \alpha }\no
&\times\Big\{\left(s_0-s_+-i\epsilon\right)^{\alpha}+\left(s_0-s_--i\epsilon\right)^{\alpha}\Big\}\co
\label{eq:T1R}
\end{align}
where $s_+$ and $s_-$ stand for $s_{\pm}=(p\pm q)^2=\mN^2\pm2\mN \nu-Q^2$ and $s_0\geq \mN^2$ is a constant. The expression~\eqref{eq:T1R} is manifestly symmetric under photon crossing. Unless the intercept $\alpha$ is an integer,\footnote{Integer values of $\alpha$ require special treatment, but since this case does not arise for the parameterizations we are working with, we do not discuss it further.} the first term in the curly brackets contains a branch cut along the positive real axis, starting at $2m\nu=s_0-m^2+Q^2$. The second is real there. One readily checks that, on the upper rim of this cut, the individual terms in the sum~\eqref{eq:T1R} differ from the asymptotic expression~\eqref{eq:T1RStandard} only through contributions of $O(s^{\alpha-1})$.

The basic assumption made in~\cite{GL1975}  is that, in the limit $\nu\rightarrow \infty$ at fixed $q^2$, only the Reggeons survive, so that the difference tends to zero:\footnote{More precisely, it is assumed that the difference disappears rapidly, so that it obeys an unsubtracted dispersion relation.}
\be
\label{eq:DeltaT1as}
T_1 (\nu,q^2)- T^R_1(\nu,q^2)\rightarrow 0\co
\ee 
We refer to this hypothesis as {\it Reggeon dominance}.  A nonzero limiting value in~\eqref{eq:DeltaT1as} would represent a $\nu$-independent term. In Regge-language, a term of this type would correspond to a fixed pole at angular momentum $J=0$. The Reggeon dominance hypothesis~\eqref{eq:DeltaT1as} thus excludes the occurrence of such a fixed pole.\esp

The presence or absence of a fixed pole at $J=0$ in Compton scattering is a standard topic in Regge pole theory~\cite{Collins:1977jy} and the literature contains several works advocating the presence of such a contribution. In particular, the universality conjecture formulated in~\cite{Brodsky:2008qu} has received considerable attention (see e.g.~\cite{Muller:2015vha} and the papers quoted therein). 

\bsp
Note, however, that these considerations go beyond the safe grounds provided by asymptotic freedom. While the short distance properties of QCD ensure that, if both $\nu$ and $q^2$ are large, the behaviour of $T_1(\nu,q^2)$ and $T_2(\nu,q^2)$ is governed by the perturbative expansion in powers of the strong coupling constant, the behaviour in the Regge region, where only $\nu$ becomes large, is not controlled by the short distance properties of QCD. In particular, values of $q^2$ of the order of $\Lambda_\text{QCD}^2$ are outside the reach of perturbation theory, even if $\nu$ is large.

The perturbative analysis shows that an infinite set of graphs needs to be summed up to understand the high-energy behaviour of the amplitudes in the Regge region. The dominating contributions can be represented in terms of poles and cuts in the angular momentum plane (Reggeon calculus, Reggeon field theory). The behaviour of the sum thus differs qualitatively from the one of the individual diagrams. 

There is solid experimental evidence for the presence of Reggeons also in the data. The relation~\eqref{eq:DeltaT1as} amounts to the assumption that the asymptotic behaviour of the current correlation function can be understood in terms of these. In the analysis described in the present paper, this assumption plays a key role. In particular,
as will be demonstrated explicitly below, it uniquely fixes the subtraction function relevant for the difference between proton and neutron  in terms of the electron cross sections, so that the entire self-energy difference can be expressed in terms of these cross sections. In other words, the necessity of a subtraction in the fixed-$q^2$ dispersion relation for $T_1(\nu,q^2)$ modifies the relation between the self-energy difference and the electron cross sections, but does not destroy it.\footnote{In~\cite{GL1975}, this conclusion was derived on the basis of a somewhat weaker form of Reggeon dominance, which does not invoke the matrix elements of the time-ordered product, but those of the current commutator. For a brief discussion of this aspect, we refer to \app~\ref{app:Causality}.}

\esp

The subtraction functions occurring in the fixed-$t$ dispersion relations relevant for {\it real} Compton scattering are analyzed in \cite{Caprini1,Caprini2}. As shown there, the experimental information on the differential cross sections can be used to impose bounds on the subtraction functions. In particular, these bounds lead to the conclusion that the electric polarizability of the proton is necessarily larger than the magnetic one, in conformity with experiment. An update of this work with the data available today is highly desirable. Unfortunately, this approach to the problem cannot readily be extended to virtual Compton scattering, because data on the differential cross sections are available only for real photons.  

\subsection{Recent work}
\label{subsec:Recent work}

\bsp
The numerical analysis of~\cite{GL1975} was based on the scaling laws proposed by Bjorken~\cite{Bjorken}. The data available at the time were perfectly consistent with these, but Bjorken scaling correctly accounts for the short-distance properties of QCD only to leading order in the perturbative expansion in powers of $\alpha_s$.  The higher-order contributions generate specific violations of Bjorken scaling~\cite{Gross:1973id,Politzer:1973fx}. In the meantime, the implications of the phenomenon and the corresponding modification of the short-distance properties of the matrix elements $\langle p|Tj^\mu(x)j^\nu(y)|p\rangle$   have been investigated by Collins~\cite{Collins}. Unfortunately, however, he did not reevaluate the self-energy difference in this framework. In fact, the question of whether the Reggeons do dominate the asymptotic behaviour or whether the amplitude in addition contains a fixed pole at $I = 1$, $J = 0$ is not touched at all in that work. 
\esp

\bsp
Motivated in part by the study of hadron electromagnetic mass shifts on the lattice
(see, e.g., \cite{Horsley:2013qka,Borsanyi:2013lga,Borsanyi:2014jba}), the Cottingham
formula has recently been reexamined~\cite{WalkerLoud:2010qq,WalkerLoud:2012bg,WalkerLoud:2012en,WalkerLoud:2013yua,WalkerLoud:2014iea,Thomas:2014dxa,Erben:2014hza}, but the central issue in this context -- the possible occurrence of fixed poles -- is not addressed in these papers, either. Instead,
the electron cross sections $\sigma_{\indT}$, $\sigma_{\indL}$ and the subtraction function $S_1(q^2)$ are treated as physically independent quantities. The main problem with the framework set up in~\cite{WalkerLoud:2012bg} is that a direct experimental determination for $S_1(q^2)$ is not available. To bridge the gap, the authors set up a model which parameterizes the dependence of the subtraction function on $q^2$. The overall normalization, $ S_1(0)$, can in principle be determined from the difference between the magnetic polarizabilities of proton and neutron, albeit the experimental value is subject to rather large uncertainties~\cite{Griesshammer}. The main problem in this approach, however, is the momentum-dependence of the subtraction function, which leads to a systematic uncertainty that is difficult to quantify.
\esp

\subsection{Structure of the present paper}
\label{subsec:Structure}

\bsp
The remaining sections are organized as follows.
In \sec~\ref{sec:Determination}, we show how the Reggeon dominance hypothesis~\eqref{eq:DeltaT1as} fixes  the subtraction function $S_1(q^2)$  from space-like data alone. In \sec~\ref{sec:Elastic}, we discuss the splitting of the amplitudes $T_i$  into elastic and inelastic contributions. 
We derive  sum rules for the nucleon polarizabilities in \sec~\ref{sec:Polarizability}, while a
thorough phenomenological analysis is provided in \sec~\ref{sec:numerical}. In particular, the sum rules 
allow us to predict the difference between the electric polarizabilities of proton and neutron. In view of the fact that the proton polarizabilities are experimentally known more accurately, our result can be turned into a prediction of the electric polarizability of the neutron, which is consistent with observation but somewhat more precise. The magnetic polarizabilities then follow from the Baldin sum rule. 
Section~\ref{sec:Self-energy} is devoted to the electromagnetic self-energies
 of proton and neutron. We discuss the renormalization of the Cottingham formula, in particular the role of the subtraction function in
 the evaluation of the self-energy and provide  a comparison with recent work on the issue. A summary and concluding remarks are given in \sec~\ref{sec:Conclusion}. In \app~\ref{app:Notation}, we detail the notation used. Appendix~\ref{app:LET} reviews those properties of Compton scattering we are making use of. In particular, we discuss the frame-dependence of the spin average and derive the low-energy theorem which underlies the sum rule for the electric polarizability. Appendices~\ref{app:Causality} and~\ref{app:New} contain a short discussion of the role of causality in our analysis. Last but not least, we note that
 in~\cite{WalkerLoud:2010qq,WalkerLoud:2012bg,WalkerLoud:2012en,WalkerLoud:2013yua},
 a comparison with the analysis of~\cite{GL1975} is attempted. Unfortunately, many of the statements made there are simply incorrect. 
Some of the misconceptions are rectified in \app~\ref{app:Comments}. 
\esp

\section{Determination of the subtraction function}
\label{sec:Determination}

The Regge amplitude obeys a once-subtracted fixed-$q^2$ dispersion relation: 
\be 
\label{eq:T1R0} 
T_1^R(\nu,q^2) =T_1^R(0,q^2)+2\nu^2\int_0^\infty \frac{d\nu'}{\nu'} \,\frac{V_1^R(\nu',q^2)}{\nu'^2-\nu^2-i\epsilon}\fs
\ee
In the space-like region and for $\nu\geq 0$, the absorptive part of the amplitude specified in~\eqref{eq:T1R} is given by
\be
\label{eq:V1R}  
V^R_1(\nu,q^2) =\sum_{\alpha>0} \beta_\alpha(Q^2)\,\theta(s_+-s_0) \left( s_+-s_0\right)^{\alpha}\fs
\ee

\bsp
The Reggeon dominance hypothesis~\eqref{eq:DeltaT1as} implies that the difference between the full amplitude and the Regge contributions,  $\overline{T}_1(\nu,q^2)\equiv  T_1(\nu,q^2)- T_1^R(\nu,q^2)$, obeys an unsubtracted dispersion relation. In particular, the value of $\overline{T}_1(0,q^2)=S_1(q^2)-T_1^R(0,q^2)$ is given by an integral over the difference $V_1(\nu,q^2)-V_1^R(\nu,q^2)$.  Hence the subtraction function can be represented as 
\begin{align}   
S_1(q^2)&=T_1^R(0,q^2)\no
&+ 2 \int_0^\infty \frac{d\nu}{\nu} \,\left\{ V_1(\nu,q^2)- V_1^R(\nu,q^2)\right\}\fs
\label{eq:S}
\end{align}
This formula explicitly represents the subtraction function in terms of measurable quantities: the structure function $ V_1(\nu,q^2)$ is determined by the cross sections for inclusive electron-nucleon scattering. The high-energy behaviour of these cross sections also determines the Reggeon residues $\beta_\alpha(Q^2)$ and thereby fixes the term $T^R_1(0,q^2)$, as well as the corresponding contribution to the structure function, $V_1^R(\nu,q^2)$.  

If the trajectory intercepts $\alpha$ were all below zero, the unsubtracted dispersion integral over $V_1^R(\nu,q^2)$ would converge and would exactly compensate the first term on the right of~\eqref{eq:S} -- the subtraction function would then be given by the unsubtracted dispersion integral over $V_1(\nu,q^2)$. The expression for the subtraction function in~\eqref{eq:S} shows how the divergence of the unsubtracted dispersion integral generated by the Reggeons is handled: the corresponding contribution is removed from the integrand, so that the integral converges also at the physical values of the intercepts. The modification is compensated by the term $T_1^R(0,q^2)$, which must be added to the integral over the remainder. The procedure amounts to analytic continuation in $\alpha$ from negative values, where $T^R_1(0,q^2)$ is given by the unsubtracted dispersion integral over $V_1^R(\nu,q^2)$ to the physical values, where that representation does not hold any more. 
\esp

We emphasize that the specific form used for the Regge parameterization does not matter. In particular, the Regge amplitude specified in~\eqref{eq:T1R} involves a free parameter, $s_0$. Since it does not affect the leading term in the asymptotic behaviour, the value used for $s_0$ is irrelevant -- our results are independent thereof. In the following, we simplify the equations by taking $s_0$ in the range $s_0\geq (\mN+\mpi)^2$, which has the advantage that $V_1^R(\nu,q^2)$ then vanishes outside the inelastic region.  

\section{Elastic and inelastic contributions}
\label{sec:Elastic}

\subsection{Elastic part}

\bsp
The contributions to the structure functions arising from the elastic reaction $e+N\rightarrow e+N$ are determined by the electromagnetic form factors of the nucleon. In the space-like region, these contributions are restricted to the lines $q^2 =\pm 2\nu \mN$ and read
($i = 1,2$)
\begin{align}
 V_i^\el(\nu,q^2)&= v_i^\el(q^2)\left\{\delta(q^2+2\mN \nu)-\delta(q^2-2\mN \nu)\right\}\co\no
v_1^\el(q^2)&=\frac{2\mN^2}{4\mN^2-q^2}\left\{G_E^2(q^2)-G_M^2(q^2)\right\}\co\no
v_2^\el(q^2)&=\frac{2\mN^2}{(-q^2)(4\mN^2-q^2)}\no
&\times\left\{4\mN^2G_E^2(q^2)-q^2G_M^2(q^2)\right\}\co
 \label{eq:Vel}
\end{align}
where $G_E(t)$ and $G_M(t)$ are the Sachs form factors. 
\esp

The elastic contributions to the time-ordered amplitudes $T_1,T_2$ cannot be specified as easily. In perturbation theory, they are usually referred to as Born terms and it is not a trivial matter to specify them at higher orders of the calculation. In effective low-energy theories, the decomposition into a Born term and a 'structure part' is not a simple matter, either. For a detailed discussion of these aspects, we refer to~\cite{Scherer:1996ux,McGovern:2000cd,Birse:2012eb}. In the framework of dispersion theory, however, the decomposition is unambiguous. The reason is that analytic functions are fully determined by their singularities and their asymptotic behaviour: dispersion theory provides a representation of the amplitudes in terms of its singularities. In our framework, this representation is given by the dispersion relations~\eqref{eq:T1} and the sum rule~\eqref{eq:S}. The elastic contribution is the part of the amplitude which is generated by the singularities due to the elastic intermediate states. These are specified in~\eqref{eq:Vel}. Accordingly, the elastic parts of $T_1,T_2$ are obtained by simply replacing $V_1,V_2$ with $V_1^\el,V_2^\el$ and dropping the Regge contributions.  In the case of $T_2$, this leads to 
\be
\label{eq:T2ela} 
T_2^\el(\nu,q^2)=2\int_0^\infty d\nu' \nu' \,\frac{V_2^\el(\nu',q^2)}{\nu'^2-\nu^2-i\epsilon}\fs
\ee
In the case of $T_1(\nu,q^2)$ there are two contributions, one from the subtraction function, the other from the subtracted dispersion integral:
\be
\label{eq:T1ela} 
T_1^\el(\nu,q^2)=S^\el_1(q^2)+2\nu^2\int_0^\infty \frac{d\nu'}{\nu'} \,\frac{V_1^\el(\nu',q^2)}{\nu'^2-\nu^2-i\epsilon}\fs
\ee
The sum rule~\eqref{eq:S} for the subtraction function implies
\be
\label{eq:T1elb} 
S^\el_1(q^2)=2 \int_0^\infty \frac{d\nu}{\nu} \,V_1^\el(\nu,q^2)\fs
\ee
Taken together, the two terms on the right hand side of~\eqref{eq:T1ela} yield the unsubtracted dispersion integral, so that the expression takes the same form as the one for $T_2^\el(\nu,q^2)$:
\be
\label{eq:T1elc} 
T_1^\el(\nu,q^2)=2\int_0^\infty d\nu' \nu'\, \frac{V_1^\el(\nu',q^2)}{\nu'^2-\nu^2-i\epsilon}\fs
\ee
Inserting the explicit expressions for the elastic contributions to the structure functions, we obtain
\begin{align}
 T_1^\el(\nu,q^2)&= \frac{4\mN^2q^2}{(4\mN^2\nu^2-q^4)(4\mN^2-q^2)}\no
 &\times\left\{G_E^2(q^2)-G_M^2(q^2)\right\}\co\no
T_2^\el(\nu,q^2)&= -\frac{4\mN^2}{(4\mN^2\nu^2-q^4)(4\mN^2-q^2)}\no
&\times\left\{4\mN^2G_E^2(q^2)-q^2G_M^2(q^2)\right\}\fs
\label{eq:Tel} 
\end{align}
Both functions tend to zero when $\nu$ becomes large: by construction, the elastic part of $T_1(\nu,q^2)$ does not contain a singularity at infinity. Moreover,
as demonstrated in \app~\ref{app:New}, even taken by itself, the elastic contributions can be represented in manifestly causal form.

The explicit expression for the elastic part of the subtraction function,
\be
\label{eq:Sel} 
S^\el_1(q^2)=-\frac{4\mN^2}{q^2(4\mN^2-q^2)}(G_E^2(q^2)-G_M^2(q^2))\co
\ee
exclusively involves the form factors, which are known very precisely.  
 
\subsection{Inelastic part}

We refer to the remainder as the inelastic part of the amplitude:
\be
\label{eq:inel}
T_i(\nu,q^2)=T_i^\el(\nu,q^2)+T_i^\inel(\nu,q^2)\co\quad i= 1,2\fs
\ee
In contrast to the elastic part, which contains the poles generated by the elastic intermediate states and is singular at the origin, the inelastic part is regular there. At high energies, the converse is true: while the elastic part tends to zero, the inelastic part includes the contributions from the Reggeons, which are singular at infinity. In particular, the sum rule for the inelastic part of the subtraction function reads:  
\begin{align}
S^\inel_1(q^2)&=  T_1^R(0,q^2)\no
&+2\int_{\nuth}^\infty\frac{d\nu}{\nu} \left\{V_1(\nu,q^2)-V_1^R(\nu,q^2)\right\} \co
\label{eq:Sinel}
\end{align}
where $\nuth=\mpi+(\mpi^2-q^2)/2\mN$ denotes the inelastic threshold. The dispersive representation for the inelastic part of $T_1(\nu,q^2)$ then becomes
\begin{align}
T_1^\inel(\nu,q^2)&=S^\inel_1(q^2)\no
&+  2\nu^2  \int_{\nuth}^\infty \frac{d\nu'}{\nu'} \,\frac{V_1(\nu',q^2)}{\nu'^2-\nu^2-i\epsilon}\fs
\label{eq:T1ineldisp}
\end{align}
In the case of $T_2(\nu,q^2)$, a subtraction is not needed. The contribution from the elastic intermediate state to the dispersion integral in~\eqref{eq:T1} coincides with the expression for $T_2^\el(\nu,q^2)$ in~\eqref{eq:Tel}. Removing this part, which is even more singular at the origin than $T_1^\el(\nu,q^2)$, we obtain the following representation for the inelastic part:
\be
\label{eq:T2inel} 
T_2^\inel(\nu,q^2)=  2\int_{\nuth}^\infty d\nu'\, \nu' \,\frac{ V_2(\nu',q^2)}{\nu'^2-\nu^2-i\epsilon}\fs
\ee 

\subsection{Subtraction function in terms of cross sections}
\label{subsec:sigma}

The structure function $V_1(\nu,q^2)$ is a linear combination of the transverse and longitudinal cross sections, see \app~\ref{app:Notation}:
\begin{align}
 V_1(\nu,q^2)&=  \frac{\mN\nu}{2\alphaem}k(\nu,Q^2)\big\{\bar{\sigma}_{\indL}(\nu,Q^2)-\sigma_{\indT}(\nu,Q^2)\big\}\co\no
\bar{\sigma}_{\indL}(\nu,Q^2)&\equiv \frac{\nu^2}{Q^2}\sigma_{\indL}(\nu,Q^2)\co\no
k(\nu,Q^2)&\equiv\frac{1}{2\pi^2} \frac{\nu -Q^2/2\mN}{\nu (\nu^2+Q^2)}\fs
\label{eq:VsigmaLT} 
\end{align}
The representation of the subtraction function thus involves integrals over the transverse and longitudinal cross sections. For $S_1^\inel(q^2)$ and $S_2^\inel(q^2)$, the following integrals are relevant: 
\begin{align}
\Sigma^{\indT}(Q^2)&= \int_{\nuth}^\infty d\nu\;
k(\nu,Q^2)\,\sigma_{\indT}(\nu,Q^2)\co\label{eq:SigmaT}\\
\Sigma^{\indL}_1(Q^2)&=
\frac{\alphaem}{\mN}T_1^R(0,q^2)\no
&+ \int_{\nuth}^\infty d\nu\; 
k(\nu,Q^2)\,\Delta\bar{\sigma}_{\indL}(\nu,Q^2)\co\label{eq:Sigma1L}\\
\Sigma^{\indL}_2(Q^2)&=   \int_{\nuth}^\infty d\nu\;
k(\nu,Q^2)\,\sigma_{\indL}(\nu,Q^2)\co \label{eq:Sigma2L}\\
\Delta\bar{\sigma}_{\indL}(\nu,Q^2)&\equiv\bar{\sigma}_{\indL}(\nu,Q^2)-\bar{\sigma}_{\indL}^R(\nu,Q^2)\fs\label{eq:Deltasigma}
\end{align}
Expressed in terms of these, $S_1^\inel(q^2)$ and $S_2^\inel(q^2)$ are given by 
\begin{align}
S_1^\inel(q^2)&=\frac{\mN}{\alphaem}\Sigma_1(Q^2)\co\no
S_2^\inel(q^2)&=\frac{\mN}{\alphaem}\Sigma_2(Q^2)\co\no
\Sigma_1(Q^2)&=-\Sigma^{\indT}(Q^2)+\Sigma^{\indL}_1(Q^2) \co\no
\Sigma_2(Q^2)&=\Sigma^{\indT}(Q^2)+\Sigma^{\indL}_2(Q^2) \fs
\label{eq:SSigma}
\end{align}
While the transverse parts of $\Sigma_1(Q^2)$ and $\Sigma_2(Q^2)$ only differ in sign, the longitudinal parts are quite different.
Regge asymptotics implies that $\sigma_{\indT}$ as well as $\sigma_{\indL}$ grow in proportion to $\nu^{\alpha-1}$. Accordingly, the integral $\Sigma^{\indT}(Q^2)$ converges -- it represents a generalization of the integral relevant for the Baldin sum rule to $Q^2\neq 0$ (cf.\ \subsec~\ref{subsec:Sum rule}). 
While $\Sigma_2^{\indL}(Q^2)$ is dominated by the contributions from the low-energy region and rapidly converges as well, it is essential that Reggeon exchange be accounted for in $\Sigma_1^{\indL}(Q^2)$. 

We are assuming that, at high energies, the longitudinal cross section can be approximated with a representation of the form
\begin{align}
 \bar{\sigma}_{\indL}^R(\nu,Q^2)&=8\pi^2\alphaem\,\frac{\nu^2+Q^2}{2\mN \nu -Q^2}\no
 &\times\sum_{\alpha>0}  \beta_\alpha(Q^2)(2\mN\nu-Q^2+\mN^2-s_0)^\alpha  \fs
 \label{eq:sigmaLR}
\end{align}
At $Q^2=0$, a Reggeon term proportional to $\nu^{\alpha}$ in $V_1$ corresponds to a contribution to $\bar{\sigma}_{\indL}$ that is proportional to $\nu^{\alpha+1}$. For nonzero values of $Q^2$, however, the factor in front of the sum implies that the corresponding cross section contains sub-leading contributions. As discussed at the end of \sec~\ref{sec:Determination}, the specific form used for the Regge parameterization is not essential -- as long as it satisfies a once-subtracted dispersion relation and  correctly represents the asymptotic behaviour of the physical cross section. We stick to the one specified in~\eqref{eq:T1R}, which leads to~\eqref{eq:sigmaLR}.

\subsection{Chiral expansion}
\label{subsec:chiPT}

Chiral perturbation theory ($\chi$PT) exploits the fact that in the limit $m_u,m_d\rightarrow 0$ (at fixed $\Lambda_\text{QCD}, m_s, ..., m_t$) QCD acquires an exact chiral symmetry, which strongly constrains the low-energy properties of the amplitudes. The chiral perturbation series provides a representation of the quantities of interest in powers of momenta and quark masses. In the chiral limit, the pion is a massless particle, but when the quark masses $m_u,m_d$ are turned on, the pion picks up mass in proportion to the square root thereof, $\mpi^2 =(m_u+m_d) B+O(m_q^2 \log m_q)$.

\bsp
In the context of the present paper, we only need the chiral expansion of the form factors $G_E(q^2)$, $G_M(q^2)$, and of the functions $S_1(q^2)$, $S_2(q^2)$.  These quantities involve a single momentum variable, $q^2$. As we work in the isospin limit $m_u=m_d$, the corresponding chiral perturbation series involves an expansion in the two variables $\mpi$ and $q^2$. The series can be ordered in powers of $\mpi$; the coefficients then depend on the ratio 
\be
\label{eq:tau}
\tau=-\frac{q^2}{4\mpi^2}\co
\ee 
which counts as a quantity of $O(1)$. In contrast to the straightforward Taylor series in powers of $q^2$, the chiral expansion is able to cope with the infrared singularities generated by the pions. 
\esp 

\bsp
To leading order in the chiral expansion, the infrared singularities are described by a set of one-loop graphs of the effective theory~\cite{Gasser:1987rb}. In the case of the magnetic Sachs form factor, for instance, the evaluation of the relevant graphs within Heavy Baryon $\chi$PT leads to the following expression for the first non-leading term in the chiral expansion~\cite{Bernard:1992qa}:\footnote{Note that the range of validity of the representation~\eqref{eq:GMchiPT} is limited. The $\pi\pi$ intermediate states generate a  branch point in the form factors at $q^2=4\mpi^2$, which corresponds to $\tau=-1$. While relativistic formulations of Baryon $\chi$PT do cover this region, an infinite series of Heavy Baryon $\chi$PT graphs contributes in the vicinity of that point, more precisely in the region where $\tau+1$ is small, of $O(\mpi^2/\mN^2)$~\cite{Becher:1999he}. In the present paper, however, we make use of the chiral expansion only near $\tau=0$, where the nonrelativistic framework is adequate.}  
\vfill

\begin{align}
 G^p_M(q^2) &=  \mu^p - \frac{g_A^2 \mN \mpi}{16\pi F_\pi^2}
\left\{(1+\tau)\frac{\arctan\sqrt{\tau}}{\sqrt{\tau}}-1\right\}\no
&+O(\mpi^2\log \mpi)\co\no
G^n_M(q^2) &=  \mu^n +\frac{g_A^2 \mN \mpi}{16\pi F_\pi^2}
\left\{(1+\tau)\frac{\arctan\sqrt{\tau}}{\sqrt{\tau}}-1\right\}\no
&+O(\mpi^2\log \mpi)\fs
\label{eq:GMchiPT} 
\end{align}
Up to and including $O(\mpi)$, the magnetic form factor can thus be represented in terms of the magnetic moment $\mu$,   the pion decay constant, $F_\pi=92.21(14)\MeV$~\cite{Rosner}, and  the nucleon matrix element of the axial charge, $g_A=1.2723(23)$~\cite{PDG}. 
The formula shows that, up to higher-order contributions, the singularity is described by a function of the ratio $\tau=(-q^2)/4\mpi^2$: the scale is set by the pion mass, not by $\Lambda_\text{QCD}$. The presence of a scale that disappears in the chiral limit also manifests itself in the slope of the form factor at $q^2=0$, i.e.\ in the magnetic radius: the above representation shows that the chiral expansion of the magnetic radii of proton and neutron starts with a term of $O(1/\mpi)$. 
\esp

The low-energy behaviour of the electric Sachs form factors is less singular: 
\begin{align} 
G^p_E(q^2)&= 1+O(\mpi^2\log \mpi)\co\no
G^n_E(q^2)&=O(\mpi^2\log \mpi)\fs
\label{eq:GEchiPT}
\end{align}
Accordingly, the chiral expansion of the electric radii does not start with a term of $O(1/\mpi)$, but with a chiral logarithm, comparable to the situation with the charge radius of the pion.

\bsp
The subtraction function also diverges if the chiral limit is taken at a fixed value of the ratio $q^2/\mpi^2$: the leading term in the chiral expansion of $S^\inel_1(q^2)$ is of order $1/\mpi$ and is determined by $F_\pi$ and $g_A$ as well~\cite{Nevado:2007dd}:
\begin{align}
S^\inel_1(q^2) &=  -\frac{g_A^2 \mN}{64\pi F_\pi^2 \mpi\tau}\left\{1-\frac{\arctan\sqrt{\tau}}{\sqrt{\tau}}\right\}\no
&+O(\log \mpi)\fs
\label{eq:S1chiPT}
\end{align}
The expansion of the analogous term in $T_2$ starts with~\cite{Nevado:2007dd}:
\begin{align}
 S_2^\inel(q^2)&=  -\frac{g_A^2 \mN}{64\pi F_\pi^2 \mpi\tau}\left\{1-(1+4\tau)\frac{\arctan\sqrt{\tau}}{\sqrt{\tau}}\right\}\no
 &+O(\log \mpi)\fs
 \label{eq:S2chiPT}
 \end{align}
In either case, the leading term is the same for proton and neutron -- for $S^\inel_1(q^2)$ and $S^\inel_2(q^2)$, the chiral expansion of the difference between proton and neutron only starts at $O(\log \mpi)$.  
\esp
 
\begin{table*}[th!]
\centering
\renewcommand{\arraystretch}{1.3}
\begin{tabular}{crcrrcrrc}
\toprule
&$\alpha_E$ &&$\alpha_E^\el$&$\beta_M$&&$\beta_M^\el$&$\alpha_E+\beta_M$\\
\midrule
$p$&$10.65(0.50)$ &\cite{McGovern:2012ew} &$0.55$&$3.15(0.50)$ & \cite{McGovern:2012ew}&$-0.55$&$13.8(4)$& \cite{Olmos}\\
$n$&$11.55(1.50)$ &\cite{Myers:2014ace} &$0.62$&$3.65(1.50)$ &\cite{Myers:2014ace}&$-0.62$&$15.2(4)$& \cite{Levchuk}\\
$p-n$&$-0.9(1.6)$&&$-0.08$&$-0.5(1.6)$&&$0.08$&$-1.4(6)$&\\
\bottomrule
\end{tabular}
\renewcommand{\arraystretch}{1.0}
\caption{Experimental values of the nucleon polarizabilities, in units of $10^{-4}\text{fm}^3$, as determined from EFT extractions in Compton scattering~\cite{McGovern:2012ew,Myers:2014ace,Myers:2015aba} and analyses of the Baldin sum rule~\cite{Olmos,Levchuk} (see also~\cite{Griesshammer,Babusci:1997ij}). The latter results were imposed in~\cite{McGovern:2012ew,Myers:2014ace}, so that the quoted errors for $\alpha_E$ and $\beta_M$ are anticorrelated.}
\label{tab:polarizabilities}
\end{table*}

\section{Nucleon polarizabilities}
\label{sec:Polarizability} 

\subsection{Low-energy theorems}
\label{subsec:Low energy theorem}

\bsp
In contrast to the elastic parts, which are singular at the origin, the inelastic contributions to $T_1(\nu,q^2), T_2(\nu,q^2)$ do admit a Taylor series expansion in powers of $\nu$ and $q^2$. Two  low-energy theorems relate the leading terms in this expansion to the polarizabilities of the nucleon.  The theorems amount to rather nontrivial statements, because the functions $T_1(\nu,q^2), T_2(\nu,q^2)$  represent the virtual Compton scattering amplitude in the forward direction, while the experimental determination of the polarizabilities relies on real Compton scattering at nonzero scattering angle. A concise derivation is given in \app~\ref{app:LET}. 
\esp

\bsp
In the above notation, the low-energy theorems take the simple form: 
\begin{align}
  S_1^\inel(0)&= - \frac{\kappa^2}{4\mN^2}-\frac{\mN}{\alphaem}\,\beta_M\co\label{eq:LET1}\\ 
 S_2^\inel(0)&=   \frac{\mN}{\alphaem}\,(\alpha_E+\beta_M)\co \label{eq:LET2}
\end{align}
where $\kappa$ is the anomalous magnetic moment, $\alpha_E$ and $\beta_M$ are the electric and magnetic polarizabilities of the particle, and $\alphaem$ is the fine structure constant. These relations show that the polarizabilities contain an elastic as well as an inelastic part, while their sum, $\alpha_E+\beta_M$, is purely inelastic:
\be 
\label{eq:betael} 
\alpha_E^\el= \frac{\alphaem\kappa^2}{4\mN^3}\co\quad
\beta_M^\el=-\frac{\alphaem\kappa^2}{4\mN^3}\fs
\ee
Table~\ref{tab:polarizabilities} shows that the elastic parts only represent a small fraction of the polarizabilities.
\esp

\subsection{Sum rules for the polarizabilities}
\label{subsec:Sum rule}

The left hand side of the low-energy theorem~\eqref{eq:LET1} represents the inelastic part of the subtraction function at $q^2=0$:  
\be
\label{eq:betainel} 
\beta_M^\inel=-\frac{\alphaem}{\mN}\,S^\inel_1(0)\fs
\ee
The representation for the subtraction function in~\eqref{eq:Sinel} thus amounts to a sum rule for the inelastic part of the magnetic polarizability. Adding the elastic contribution, the sum rule takes the form
\be
\label{eq:betaSigma}  
\beta_M = \Sigma^{\indT} (0)-\Sigma^{\indL}_1(0)-\frac{\alphaem\kappa^2}{4\mN^3}\fs
\ee
To our knowledge, this sum rule is new. It states that, in the absence of fixed poles, the magnetic polarizabilities of proton and neutron are determined by the cross sections for photo- and electroproduction. If the amplitude $T_1(\nu,q^2)$ were to obey an unsubtracted dispersion relation, the Regge terms in the expression for $\Sigma^{\indL}_1(Q^2)$ could be dropped, so that the sum rule would reduce to the one proposed in~\cite{Gorchtein:2008aa}. Regge asymptotics implies that a subtraction is needed,  but if the Reggeon trajectories and residues are known, the subtraction can be expressed in terms of these.  

Evaluating the dispersive representation~\eqref{eq:T2inel} at $\nu=q^2=0$, we obtain
\begin{align}
S_2^\inel(0)&= 2\int_{\nuth}^\infty\frac{d\nu}{\nu}V_2(\nu,0)\no
&=\frac{\mN}{2\pi^2\alphaem}\int_{\nuth}^\infty\frac{d\nu}{\nu^2}\,\sigma_\text{tot}(\nu)
\label{eq:Baldin} 
\fs
\end{align}
The low-energy theorem~\eqref{eq:LET2} thus represents the familiar Baldin sum rule~\cite{Baldin}. 
The integral occurring here is a limiting case of the quantity $\Sigma^{\indT}(Q^2)$ introduced in~\eqref{eq:SigmaT}: the Baldin sum rule amounts to 
\be
\label{eq:BaldinSigma}
\alpha_E+\beta_M=\Sigma^{\indT} (0)\fs
\ee
Comparison with~\eqref{eq:betaSigma} shows that the electric polarizability obeys a sum rule that exclusively involves the longitudinal cross section and the anomalous magnetic moment:
\be
\label{eq:alphaSigma}
\alpha_E = \Sigma^{\indL}_1(0)+\frac{\alphaem\kappa^2}{4\mN^3} \fs
\ee

\begin{figure*}[t]
 \includegraphics[width=0.48\linewidth,clip]{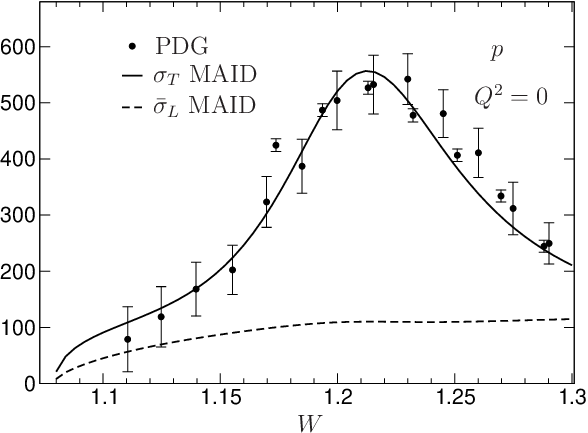}\quad
 \includegraphics[width=0.48\linewidth,clip]{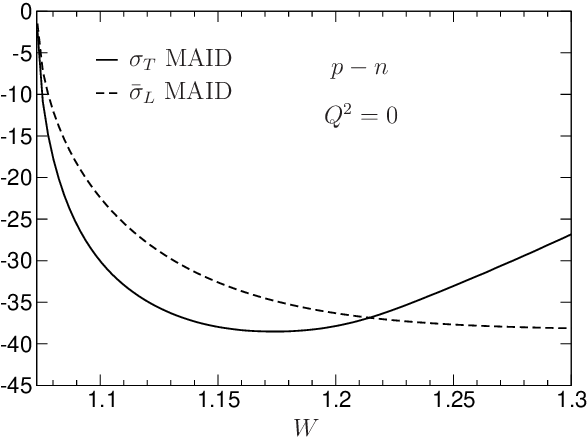}
\caption{Cross sections $\sigma_{\indT}$ and $\bar{\sigma}_{\indL}\equiv \nu^2/Q^2\sigma_{\indL}$  in the $\Delta$-region, for $Q^2= 0$.}
\label{fig:sigmaversusW}
\end{figure*}

\section{Numerical analysis}
\label{sec:numerical}

\subsection{Experimental information}
\label{subsec:exp}

We evaluate the cross section integrals on the following basis.\footnote{Throughout, the numerical values of $W$ and $Q$ refer to GeV units, the cross sections are given in $\mu$b, while the polarizabilities as well as the cross section integrals $\Sigma$ are expressed  in units of $10^{-4}\,\text{fm}^3$.}      

\bsp
\underline{\boldmath $W<1.3$}. At low energies, the resonance $\Delta(1232)$ generates the most important inelastic contribution. It decays almost exclusively into $\pi N$ final states which have been thoroughly explored. The SAID, MAID, Dubna--Mainz--Taipei (DMT), and chiral-MAID collaborations provide pion photo- and electroproduction cross sections into these channels~\cite{Drechsel:2007if,MAID,Kamalov:2000en,Hilt:2013fda,Hilt:2013coa,Workman:2012jf,SAID,DMT,chiralMAID}.\footnote{The photoproduction cross sections are also provided by the Bonn--Gatchina collaboration~\cite{Gutz:2014wit,BG}.} For 
$W< 1.3$ and real photons ($Q^2=0$), the transverse cross section is well approximated by the sum over these contributions. In particular, the representations we are using are consistent with isospin symmetry, which implies that  the contributions from the $\Delta$ to the proton and neutron cross sections are the same up to symmetry-breaking effects of $O(m_u-m_d,\alphaem)$ which are expected to be very small. 
Moreover, as seen from \fig~\ref{fig:sigmaversusW} (left panel), the $\Delta$ dominates in the
transverse cross sections and gives very small contributions to the longitudinal
ones. This property is directly related to the smallness of the $C2$ Coulomb
quadrupole form factor for the $\Delta N\gamma^*$ transition. In the non-relativistic quark model, where both the nucleon and the $\Delta$ are zero-orbital-momentum three-quark states, this form factor, as well as the one of the $E2$ electric quadrupole, vanish altogether~\cite{Becchi:1965zz,Idilbi:2003wj,Pascalutsa:2006up}. 
\esp

\begin{figure*}[t]
\centering
 \includegraphics[width=0.48\linewidth,clip]{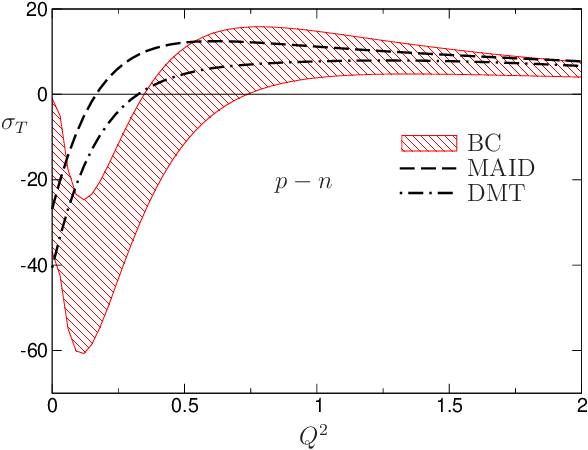}\quad 
 \includegraphics[width=0.48\linewidth]{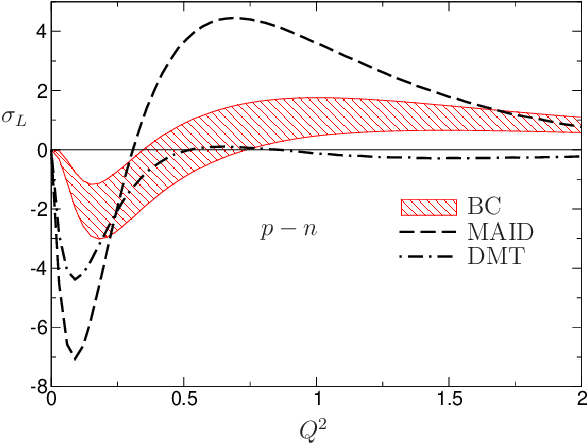}
\caption{Consistency check at the transition point $W=1.3$. The plot compares the representations of MAID and DMT used below that point with the BC-parametrization used above it. The difference between MAID and DMT and the band attached to BC represent an estimate of the uncertainties to be attached to these parameterizations. As discussed in the text, the picture implies that these parameterizations provide a coherent framework only for $Q^2>0.5$. }
\label{fig:sigmaW1300pn}
\end{figure*}

\bsp
The comparison of the full lines in the two panels of \fig~\ref{fig:sigmaversusW} shows that, in the region where the $\Delta$ generates the dominant contribution, the transverse cross sections for proton and neutron are indeed nearly the same: the differences are smaller than the individual terms by an entire order of magnitude~\cite{Becchi:1965zz}.  For the polarizabilities, the behaviour of the ratio $\sigma_{\indL}/Q^2$ in the limit $Q^2\rightarrow 0$ is relevant. Since MAID and DMT offer a representation also for this quantity, these parameterizations are particularly convenient for us. For definiteness, we identify the central values of the cross sections in the region $W<1.3$ with the average of MAID and DMT, abbreviated as MD: $\sigma_\text{MD}=\frac{1}{2}(\sigma_\text{MAID}+\sigma_\text{DMT})$.  As far as the proton cross sections are concerned, the results obtained with SAID, MAID, and DMT are practically the same, but \fig~\ref{fig:sigmaW1300pn} shows that for the small differences between proton and neutron, this is not the case. The uncertainties in the input used for the cross sections do affect our numerical results and will be discussed together with these.  
\esp

\underline{\boldmath $1.3<W<3$.} In the intermediate region, we rely on the work of Bosted and Christy (BC), who provide  parameterizations of the transverse and longitudinal proton and neutron cross sections in the resonance region, 
$\mN+\mpi<W<3.2$, in the range $0<Q^2<8$~\cite{Bosted1,Bosted2}. These contain a wealth of information, but suffer from a number of shortcomings. In particular, their fit to the data is carried out under the assumption that the ratio $\sigma_{\indL}/\sigma_{\indT}$ is
the same for proton and neutron. An experimental analysis that does not rely on this assumption would be most welcome. Second, the parameterization does not properly cover the region of very small photon virtualities (cf.~\cite{Hall:2014lea}): (a) The algebraic form of the representation used for $\sigma_{\indL}$ implies that the quantity  $\bar{\sigma}_{\indL}\equiv\sigma_{\indL}\nu^2/Q^2$ disappears when $Q$ tends to zero instead of approaching a nonzero limiting value. (b) Isospin symmetry implies that the contributions of the resonance $\Delta(1232)$ to proton and neutron are the same, but, as noted in~\cite{Erben:2014hza}, the BC-parameterization does not 
respect this symmetry to the expected accuracy. (c) The parameterization of the contribution from the resonance $N(1530)$ exhibits an unphysical  dependence on $Q^2$: in the tiny interval $0< Q^2< 0.001$, the contribution from this resonance to the transverse cross section of the proton varies by about 40\%. Although this artefact only manifests itself at very small values of $Q^2$, it seriously affects our calculation because the results obtained for the polarizabilities depend on whether we simply evaluate the sum rules at $Q^2=0$ or use very small positive values of $Q^2$ -- for the physical cross sections, a difference of this sort cannot arise. 

In the interval $1.3<W<3$, we use the following crude estimate for mean values and errors: (i) The central value is identified with the result obtained with the BC-parameterization. (ii) In order to wash out the spikes occurring at very small values of $Q^2$, we assign an 8\% uncertainty to the BC-representation of the proton cross sections: $\Delta\sigma^p = 0.08\,\sigma^p$. (iii) Since  the difference between the proton and neutron cross sections is much smaller than the individual terms, small relative errors in the latter can generate large relative errors in the difference. For this reason, we use the same error estimate for $\sigma^{p-n}$ as for the individual terms, i.e.\ work with $\Delta\sigma^{p-n}= 0.08\,\sigma^p$. 

The comparison of the representations for the difference between the proton and neutron cross sections used below and above $W = 1.3$ offers a consistency test on our calculations. Figure~\ref{fig:sigmaW1300pn} compares the representations of MAID and DMT with the uncertainty band attached to BC at the transition point. The left panel shows that the representations for the difference of the transverse cross sections used below and above that point agree with one another only for $Q^2>0.5$. The problem arises from the deficiencies mentioned above, which prevent us from reliably evaluating the cross section integrals at low values of $Q^2$. The right panel shows that the uncertainties in the difference of the longitudinal cross sections are considerable, but within these, the representations used are coherent. 

\underline{\boldmath $W>3$.} We estimate the contributions from higher energies with the representation of Alwall and Ingelman~\cite{GVMD}. It is based on the vector-meson-dominance model of Sakurai, Schildknecht, Donnachie, and Landshoff~\cite{Sakurai:1972wk,Sakurai:1973tf,Sakurai:1972zs,Donnachie} and offers a parameterization of the transverse and longitudinal cross sections of the form 
\begin{align}
 \sigma_{\indT} &= \beta^{\indT}_{\indP}(Q^2)s^{\alpha_P-1}+\beta_R^{\indT}(Q^2) s^{\alpha_R-1}\co\no
\sigma_{\indL}&= \beta^{\indL}_P(Q^2)s^{\alpha_P-1}+\beta_R^{\indL}(Q^2) s^{\alpha_R-1}\co
\label{eq:AI} 
\end{align}
where $s=W^2$ is the square of the center-of-mass energy. The Pomeron cut is approximated by a Regge pole at $\alpha_P=1.091$, while  the Reggeons with the quantum numbers of $f$ and $a_2$ are lumped together in a single contribution with $\alpha_R= 0.55$.  

The Pomeron residues of proton and neutron are the same:
\be
\label{eq:betaP} 
\beta^{\indT}_P(Q^2)^n= \beta^{\indT}_P(Q^2)^p\co\quad
\beta^{\indL}_P(Q^2)^n= \beta^{\indL}_P(Q^2)^p\fs
\ee
For the remainder, we follow~\cite{GL1975}, invoke $SU(3)$, and stick to the value of the $D/F$ ratio quoted there (for the definition of the Regge couplings $D$ and $F$ and a review of their determination, we refer to~\cite{Pilkuhn:1973wq}):
\begin{align}
 \beta_R^{\indT}(Q^2)^n&=\xi \,\beta_R^{\indT}(Q^2)^p\co\quad
 \beta_R^{\indL}(Q^2)^n=\xi\, \beta_R^{\indL}(Q^2)^p\co\no
\xi&=\frac{6F-4D}{9F-D}\simeq 0.74\fs
\label{eq:betarho}
\end{align}
\begin{figure*}[t]
\centering
 \includegraphics[width=0.48\linewidth]{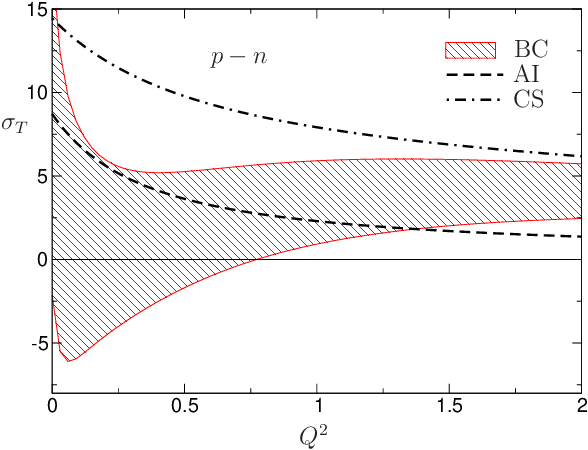}\quad
 \includegraphics[width=0.48\linewidth]{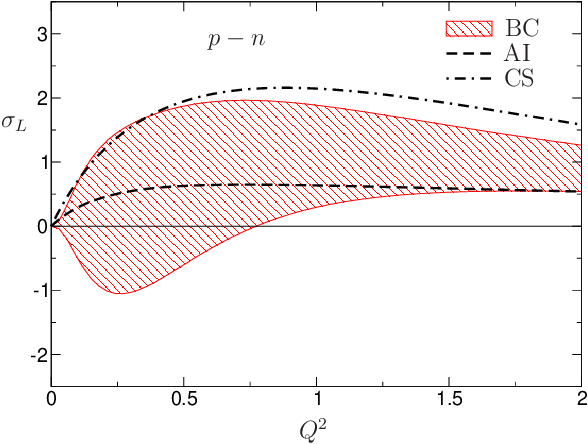}
\caption{Consistency check at the transition point $W = 3$, see main text.}
\label{fig:sigmaW3000pn}
\end{figure*}

The parameterizations for the structure function $F_2$ of Capella et al.~\cite{Capella:1994cr,Kaidalov:1998pn} and for the ratio $\sigma_{\indL}/\sigma_{\indT}$ of Sibirtsev et al.~\cite{Sibirtsev} provide an alternative Regge representation of the cross sections, which we refer to as CS. 
In figure~\ref{fig:sigmaW3000pn}, the  consistency check made at the transition point $W=1.3$ is repeated for $W=3$. The plot shows that, for $Q^2<1.4$, the central representation of AI is indeed contained in the uncertainty band attached to BC, while the one of CS runs above it. The comparison indicates that, at low values of $Q^2$, working with AI yields a coherent picture, while with CS this is not the case. For $Q^2>2$, however, the situation is reversed: there, the AI-representation yields values for the difference between the transverse cross sections that are too small while the CS-representation is consistent with the values obtained from BC. This confirms the conclusion reached in~\cite{GVMD}: the above form of the AI-representation applies as it stands only for $Q^2 < 1$. At higher values of $Q^2$, the parameterization underestimates the size of the structure function $F_2$ and further contributions have to be added for the vector-meson-dominance formulae to become compatible with the observed behaviour.  Since we do not account for these and the uncertainties we attach to the central representation do not cover the gap, the input we are working with becomes incoherent for $Q^2>2$. The right panel, on the other hand, shows that the representations we are using for the longitudinal cross section do survive the consistency test, irrespective of the value of $Q^2$.

Note that we are discussing the properties of the difference between the proton and neutron cross sections. The main problem here is that all of the well-established properties of the proton cross sections drop out when taking the difference between proton and neutron. High precision is required to measure the remainder, in particular also at high energies, where the Pomeron dominates the scenery. Also, since the longitudinal cross section is significantly smaller than the transverse one, pinning it down accurately is notoriously difficult. In both of the above representations, the ratio $\sigma_{\indL}/\sigma_{\indT}$ is taken to be energy-independent.\footnote{For the representation~\eqref{eq:AI}, this property implies $\beta_{\indR}^{\indL}/\beta_{\indR}^{\indT}=\beta_{\indP}^{\indL}/\beta_{\indP}^{\indT}$. In the notation of~\cite{GVMD}, it corresponds to $B_V/A_ V=B_\gamma/A_\gamma$, for $V=\rho,\omega,\phi$.} This appears to be consistent with experiment, but since we are not aware of a theoretical explanation, a test of the validity of this assumption would be very useful. For recent applications of these representations to the amplitudes under consideration we refer to~\cite{Hall,Sibirtsev:2013cga,Hall:2014lea}. 

\subsection{Evaluation of $\Sigma^{\indT}$ and $\Sigma_2$ for the proton}

\begin{figure}[t]
\centering
\includegraphics[width=\linewidth]{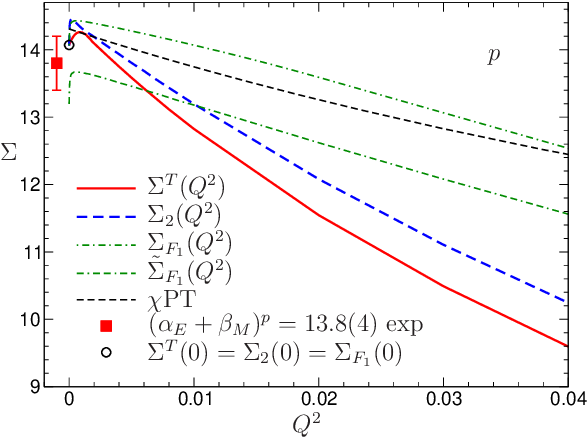} 
\caption{Cross section integrals related to the Baldin sum rule for the proton (numerical values in units of $10^{-4}\text{fm}^3$).  At $Q^2=0$, all of the quantities shown reduce to $\alpha_E^p+\beta_M^p$. The plot focuses on small values of $Q^2$, where the behaviour is dominated by the Nambu--Goldstone bosons -- in the chiral limit these generate an infrared singularity.  The short-dashed line shows the parameter-free result obtained from $\chi$PT at leading order. The cross section integrals are evaluated with the parameterization specified in \subsec~\ref{subsec:exp}, except for $\tilde{\Sigma}_{F_1}$, where the contribution from the region of the Delta is calculated with BC instead of MD.}
\label{fig:Baldin}
\end{figure}

We start the discussion of the cross section integrals with the one over the transverse cross section, $\Sigma^{\indT}(Q^2)$, which is specified in~\eqref{eq:SigmaT}. The value at the origin is relevant for the sum of the electric and magnetic polarizabilities, $\Sigma^{\indT}(0)=\alpha_E+\beta_M$.  Since the longitudinal cross section vanishes at $Q^2=0$, the function $\Sigma_2(Q^2)= \Sigma^{\indT}(Q^2)+\Sigma_2^{\indL}(Q^2)$ takes the same value there. In fact, \fig~\ref{fig:Baldin} shows that $\Sigma_2(Q^2)$ is dominated by the transverse part also at nonzero virtuality -- the longitudinal part amounts to a modest correction.  

\bsp
As pointed out in~\cite{Hall:2014lea}, the structure function $F_1(x,Q^2)$ can also be used to continue the integral relevant for the Baldin sum rule to nonzero values of $Q^2$:
\begin{align}
 \Sigma_{F_1}(Q^2)&=\frac{8\mN\alphaem}{Q^4}\int_0^{x_\text{th}}dx\, x F_1\no
 &= \frac{1}{2\pi^2 }\int_{\nuth}^\infty\frac{d\nu}{\nu^3}\;
(\nu-Q^2/2\mN)\,\sigma_{\indT} \fs
\label{eq:SigmaF1}
\end{align}
Since the integrand differs from the one relevant for $\Sigma^{\indT}(Q^2)$ only by the factor $1+Q^2/\nu^2$, the quantity $\Sigma_{F_1}$ also reduces to $\alpha_E+\beta_M$ when $Q^2$ vanishes, but drops off somewhat less rapidly when $Q^2$ grows. 
\esp

\bsp
The lines for $\Sigma^{\indT}$, $\Sigma_2$, and $\Sigma_{F_1}$ in \fig~\ref{fig:Baldin} are obtained by using the parameterizations specified in \subsec~\ref{subsec:exp}. As stated there, the contributions from the region $W<1.3$ are evaluated with the mean of MAID and DMT,  but we could just as well have used SAID -- on this plot, the difference would barely be visible. 
\esp

In~\cite{Hall:2014lea}, the function $\Sigma_{F_1}(Q^2)$ is instead evaluated with the BC-parameterization, also in the region of the $\Delta$-resonance. This leads to the behaviour indicated by the dash-dotted line labelled $\tilde{\Sigma}_{F_1}$. The topmost line, which is obtained by evaluating the same formula with the MD-parameterization, is higher by about 0.8 units. The difference is closely related to the fact that the BC-parameterization does not respect isospin symmetry to the expected accuracy (see the discussion in \subsec~\ref{subsec:SigmaT}). 

\bsp
As pointed out by Bernard, Kaiser, and Mei\ss ner~\cite{Bernard:1991rq}, $\chi$PT neatly explains the size of the combination of polarizabilities occurring in the Baldin sum rule. The parameter-free expression~\eqref{eq:S2chiPT} for the leading term in the chiral perturbation series of $\Sigma_2(Q^2)$ is shown as a dashed line. The comparison with the experimental result for $\alpha_E+\beta_M$ shows that, at small values of $Q^2$, the leading term of the chiral series dominates. In the limit $Q^2\rightarrow 0$, this term reduces to
\be
\label{eq:BaldinchiPT}
\alpha_E+\beta_M=\frac{11\alphaem g_A^2}{192 \pi F_\pi^2 \mpi}\fs
\ee
In the chiral limit this formula diverges in inverse proportion to $\mpi$: if the quarks are taken massless, $T_2^\inel(\nu,q^2)$ contains an infrared singularity at $\nu=q^2=0$.

The same singularity also shows up in the $Q^2$-dependence, which exhibits the presence of an unusually small scale: at leading order of the chiral expansion, the function $\Sigma_2(Q^2)$ depends on $Q^2$ only via the variable $\tau=Q^2/4\mpi^2$. Hence the scale is set by $2\mpi$ rather than $M_\rho$. Figure~\ref{fig:Baldin} shows that, in reality, $\Sigma_2(Q^2)$ drops even more rapidly, partly on account of the second-sheet pole associated with the $\Delta$, partly due to other higher-order contributions of the chiral series~\cite{Nevado:2007dd,Birse:2012eb,Alarcon:2013cba,Peset:2014jxa}. 
\esp

The spike seen in \fig~\ref{fig:Baldin} at tiny values of $Q^2$ illustrates the artefact mentioned in \subsec~\ref{subsec:exp}, which concerns the contribution from the resonance $N(1530)$: if the numerical values of the integrals in the region $0.002 < Q^2 <0.005$ are fit with a low order polynomial, the extrapolation to $Q^2=0$ is higher than the result of the direct evaluation at $Q^2=0$, by about 0.4 units. Since the experimental information from real Compton scattering and from photoproduction is more stringent than the one from electron scattering, which for these very small values of $Q^2$ necessarily involves extrapolations, we think that the results obtained by evaluating the integral over the transverse cross section at $Q^2=0$ are more reliable. The value obtained there with MAID or DMT is $(\alpha_E+\beta_M)^p=14.1$, while SAID yields a result that is lower by about 0.1 units. The numbers obtained at $Q^2=0$  with the parameterizations we are using thus agree with the result $(\alpha_E+\beta_M)^p=13.8(4)$ quoted in the review~\cite{Griesshammer}, which stems from~\cite{Olmos}.

\subsection{$\Sigma^{\indT}$ and $\Sigma_2$: proton-neutron difference}
\label{subsec:SigmaT}

\begin{figure}[t]
\centering
 \includegraphics[width=\linewidth]{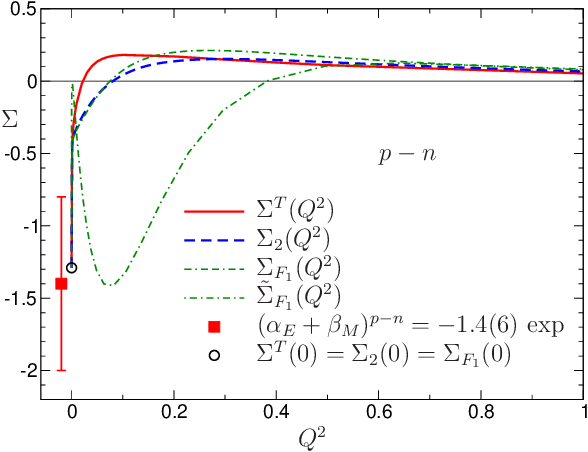}
\caption{Cross section integrals relevant for the difference between proton and neutron.}
\label{fig:SigmaTpn}
\end{figure}

\bsp
Figure~\ref{fig:SigmaTpn} shows the difference between the integrals over the proton and neutron cross sections. The picture looks very different from  \fig~\ref{fig:Baldin}: while there, the curves start at $\Sigma\simeq$ 14 and rapidly drop with $Q^2$, those in \fig~\ref{fig:SigmaTpn} start at $\Sigma\simeq$ 0 and stay there. Since the integrals under consideration are rapidly convergent, the behaviour of the cross sections in the resonance region is relevant. The reason why not much is left in the difference between proton and neutron is that, in that region, the proton and neutron cross sections are nearly the same. In particular, as mentioned in \subsec~\ref{subsec:exp}, isospin symmetry implies that the most prominent low-energy phenomenon, the $\Delta$, drops out when taking the difference between the proton and neutron cross sections. The cancellation of the main contributions also manifests itself in the chiral perturbation series: the leading terms in $\Sigma_2^p$ and $\Sigma_2^n$ are large, of order $1/\mpi$, but the coefficients are the same, so that the chiral expansion of $\Sigma_2^{p-n}$ only starts at $O(1)$.
\esp

For our cross section integrals to exhibit these features, it is essential that the representations we are using in the region of the $\Delta$ respect isospin symmetry. The dash-dotted line illustrates the fact that the BC-parameterization of the cross sections violates this constraint quite strongly: the bump seen around $Q^2\simeq 0.1$ arises from the difference between proton and neutron which occurs in that parameterization in the region of the $\Delta$. As mentioned above, the difference between the parameterizations MD and BC in the region $W<1.3$ also shows up in \fig~\ref{fig:Baldin}. It so happens that the difference between the results obtained via extrapolation from $Q^2>0.002$ and via evaluation at $Q^2=0$ nearly cancels the one between the contributions from the region of the $\Delta$ obtained with BC and with MD, so that the number obtained for $\alpha_E^p+\beta_M^p$ in~\cite{Hall} agrees with experiment.  

The spike seen at very small values of $Q^2$ is about twice as large as the one in \fig~\ref{fig:Baldin} and manifests itself much more prominently because the difference between proton and neutron is an order of magnitude smaller than the individual terms. The value obtained at $Q^2=0$ is consistent with the experimental result, $(\alpha_E+\beta_M)^{p-n}=-1.4(6)$.
 
 \subsection{Pomeron exchange}
\label{sec:Pomeron}

The integrals considered in the preceding two subsections converge rapidly. Their properties are governed by the low-energy behaviour of the cross sections -- the asymptotic behaviour does not play a significant role. For the integral $\Sigma_1^{\indL}(Q^2)$ specified in~\eqref{eq:Sigma1L}, the situation is very different: for this integral to converge, it is essential that the asymptotic behaviour of the longitudinal cross section be known, so that it can properly be accounted for. At high energies, the leading contribution stems from Pomeron exchange, which generates a branch point at $J=1$ in the angular momentum plane. In phenomenological parameterizations, such as the one specified in~\eqref{eq:AI}, the branch cut is often approximated by a Regge pole in the range $1<\alpha_P<2$.  For this parameterization to have the required asymptotic accuracy, it must describe the contribution from the Pomeron up to terms that disappear in the limit $\nu\rightarrow\infty$. 

The Regge representation we are using to describe the asymptotic behaviour of the structure functions leads to the parameterization~\eqref{eq:sigmaLR}. In this framework, the Pomeron term  in~\eqref{eq:AI} not only generates a leading contribution to the cross section with $\alpha=\alpha_P$, but also a daughter with $\alpha=\alpha_P-1$.  Furthermore, in contrast to the situation with the parameterization of the contributions from the nonleading Reggeons, the value of the parameter $s_0$ does matter here: a change in the value of $s_0$ generates an asymptotic contribution proportional to $\nu^{\alpha_P-1}$.  If the integral in~\eqref{eq:Sigma1L} does converge for one particular value of $s_0$, it diverges for any other value.

As an illustration of the mathematical problem we are facing here, consider a contribution of the form
\begin{align}
 \Delta T_1(\nu,q^2)&=\frac{1}{2}\xi(q^2)\Big\{(s_1-m^2-2m\nu-q^2)^\delta\no
 &\qquad+(s_1-m^2+2m\nu-q^2)^\delta\Big\}\co
 \label{eq:TR1}
\end{align}
which is free of fixed poles. For $\nu \geq 0$, $q^2\leq 0$, the corresponding absorptive part is given by:
\begin{align}
 \Delta V_1(\nu,q^2)&=-\frac{\sin \pi\delta}{2\pi}\xi(q^2)\,\theta(m^2+2m\nu+q^2-s_1)\no
  &\times(m^2+2m\nu+q^2-s_1)^\delta\fs
\end{align}
In the limit $\delta\rightarrow 0$, the modification of the structure function disappears, while the change in the time-ordered amplitude does not, but takes the form of a fixed-pole contribution, $\Delta T_1(\nu,q^2)\rightarrow\xi(q^2)$, which can have any desired value.

\bsp
In short: although the hypothesis that the Reggeons properly account for the behaviour at large values of $\nu$ uniquely determines the subtraction function even if Pomeron exchange contributes, the evaluation of~\eqref{eq:Sigma1L} requires knowledge of the asymptotic behaviour to an accuracy that is beyond reach. In the absence of theoretical information about the properties of the Pomeron, we are dealing with what Hadamard~\cite{Hadamard} called an ill-posed problem: in principle, the data do determine the solution, but tiny changes in the data (structure function) can lead to substantial changes in the solution (subtraction function). For this reason, we do not discuss the sum rules for the individual polarizabilities of proton and neutron any further.
\esp

A model-independent determination of the subtraction function occurring in the dispersive representation of the proton Compton amplitude is also of interest in connection with the proton radius puzzle (for a recent review see~\cite{Carlson:2015jba}). As pointed out in~\cite{Miller:2012ne}, at least part of the discrepancy could be explained if for some reason the contribution to the Lamb shift that is governed by the virtual Compton scattering amplitude were significantly larger than expected. The $\chi$PT analyses~\cite{Nevado:2007dd,Birse:2012eb,Alarcon:2013cba,Peset:2014jxa} as well as the recent works on effective field theory~\cite{Hill:2011wy} and finite-energy sum rules~\cite{Gorchtein:2013yga} were largely motivated by this puzzle; an improved  knowledge of the subtraction function would be of interest also in that context. Unfortunately, however, a major breakthrough in the theoretical understanding of the Pomeron is required before the sum rule set up above could reliably be evaluated.
  
\begin{figure}[t]
\centering
 \includegraphics[width=\linewidth]{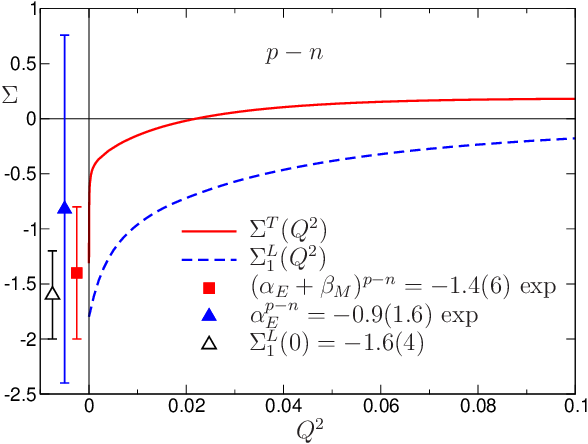}
\caption{Cross section integrals relevant for the subtraction function.}
\label{fig:Sigma1pn}
\end{figure}

\subsection[Evaluation of $\Sigma_1^{\indL}$ for proton-neutron difference]{Evaluation of $\Sigma_1^{\indL}$ for the proton-neutron difference}
\label{sec:Sigma1L}

\bsp
In the present \subsec, we focus on the difference between proton and neutron, where the Pomeron drops out: the asymptotic behaviour of $\sigma_{\indL}(\nu,Q^2)^{p-n}$ is dominated by the nonleading terms in \eqref{eq:AI}, which grow less rapidly with $\nu$, so that the problems discussed in the preceding \subsec~do not arise. The analysis of the difference is of interest for two reasons: (1) the sum rule is obtained under the same premises (absence of fixed poles, Reggeon dominance hypothesis) as the Cottingham formula. Consequently, confronting the result with existing experimental information on the polarizabilities, one may test the validity of this hypothesis. (2) our result for $\alpha_E^{p-n}$ is somewhat more accurate than the determination based on the current experimental information. Combined with the experimental values of the polarizabilities of the proton and the Baldin sum rule, this yields an improved prediction for the polarizabilities of the neutron. 
\esp

Figure~\ref{fig:Sigma1pn} compares the integrals over the transverse and longitudinal cross sections, for the difference between proton and neutron. The function $\Sigma^{\indT}(Q^2)^{p-n}$ already occurred in \fig~\ref{fig:SigmaTpn} -- we are now merely focusing on a smaller range in the variable $Q^2$. The plot shows that the integral $\Sigma_1^{\indL}(Q^2)^{p-n}$ behaves in a qualitatively different way. Both integrals are small, but while $\Sigma^{\indT}(Q^2)$ exhibits the pronounced spike at $Q^2=0$ discussed earlier, the dependence on $Q^2$ of $\Sigma_1^{\indL}(Q^2)$ is dominated by the contribution from the region of the $\Delta$, which is well understood -- in particular, the MAID and DMT representations show nearly the same $Q^2$-dependence. Using the mean of the two as central value and half of the difference as an estimate for the uncertainty for the contributions from $W<1.3$ would in our opinion represent a fair recipe, but to stay on the conservative side, we double the error estimate. For the value of the integral at $Q^2=0$ this prescription yields $\Sigma_1^{\indL}(0)_\text{MD}=-1.4(4)$. The contributions from intermediate energies, $1.3<W<3$, are small: the estimate $\Sigma_1^{\indL}(0)_\text{BC}=0.2(2)$ covers the deficiencies of the representation used there. Above that range, we use the AI-representation, attach an uncertainty of 30\% to it, and get  $\Sigma_1^{\indL}(0)_\text{AI}=- 0.3(1)$. Adding errors in quadrature, we finally obtain
\be
\label{eq:Sigma1L0}
\Sigma_1^{\indL}(0)^{p-n}=-1.6(4) \fs
\ee  

\subsection{Prediction for the polarizabilities of the neutron}
\label{sec:betan}

\bsp
In view of the relation~\eqref{eq:alphaSigma}, the result~\eqref{eq:Sigma1L0} amounts to a prediction for the difference between the electric polarizabilities of proton and neutron:
\be
\label{eq:alphapn}
\alpha_E^{p-n}=-1.7(4)\fs 
\ee
This is consistent with the current experimental value, $\alpha_E^{p-n}=-0.9(1.6)$, but significantly more precise.
The numerical result obtained from the Baldin sum rule for the difference in the value of $\alpha_E+\beta_M$ between proton and neutron, $(\alpha_E+\beta_M)^{p-n}=-1.4(6)$, then implies
\be
\label{eq:betapn}
\beta_M^{p-n}=0.3(7)\fs
\ee
According to~\eqref{eq:betainel}, this result also determines the value of the subtraction function relevant for the self-energy difference at $Q^2=0$:
\be 
\label{eq:Sinel0} 
S_1^\inel(0)^{p-n}=-0.3(1.2)\GeV^{-2}
\fs
\ee
Finally, combining the current experimental result for the electric and magnetic polarizabilities of the proton, $\alpha_E^p=10.65(50)$ and $\beta_M^p=3.15(50)$, with the numbers for $\alpha_E^{p-n}$ and $(\alpha_E+\beta_M)^{p-n}$, we arrive at a prediction for the electric and magnetic polarizabilities of the neutron:   
\be
\label{eq:alphanbetan}
\alpha_E^n=12.3(7)\co\quad\beta_M^n=2.9(0.9)\fs
\ee
These are also consistent with the current experimental values, $\alpha_M^n=11.55(1.5)$, $\beta_E^n=3.65(1.50)$, and  more precise. 
\esp
 
Note that the procedure used avoids relying on the available parameterizations of the transverse cross section, which contain sharp spikes at very small values of $Q^2$ that make the evaluation of $\Sigma^{\indT}(0)$ problematic. We make use of the fact that those present in the longitudinal cross section are much milder and allow us to assign a meaningful uncertainty to $\Sigma_1^{\indL}(0)$. We also emphasize that the fluctuations exclusively affect the behaviour at small values of $Q^2$. For the evaluation of the electromagnetic self-energy to be discussed in \sec~\ref{sec:Self-energy}, these deficiencies are of no concern, because phase space suppresses the contributions from the vicinity of the point $Q^2=0$.

\subsection{Result for the subtraction function}
\label{sec:subtraction function}

\begin{figure}[t]
\includegraphics[width=\linewidth]{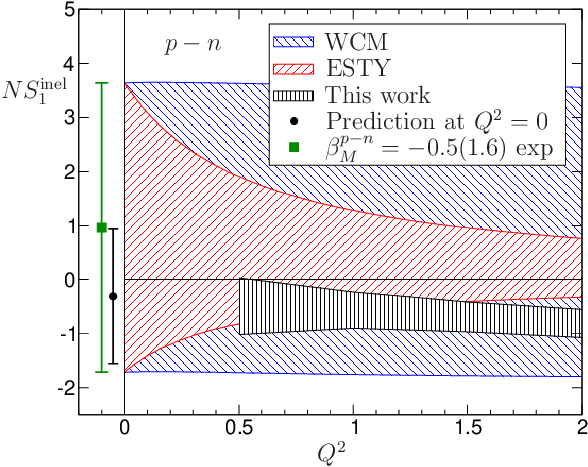}
\caption{Momentum dependence of the subtraction function (GeV units). For better visibility, the vertical axis is stretched with the inverse of the dipole form factor, $N=(1+Q^2/M_d^2)^2$, $M_d^2=0.71\GeV^2$. In this normalization, the ansatz proposed in~\cite{WalkerLoud:2012bg} (WCM) represents a broad band of nearly constant width, determined by the experimental value of the difference between the magnetic polarizabilities of proton and neutron. The curves are drawn for the current experimental value, which is indicated by the error bar on the left and concerns the value at $Q^2=0$, but is displaced to make it visible. The range obtained with the model in~\cite{Erben:2014hza} (ESTY) starts with the same width at $Q^2=0$, but shrinks as $Q^2$ grows. 
The comparatively narrow third band represents our work. We do not present an error estimate in the region $0<Q^2<0.5$, because there, our results are sensitive to the inadequacies of the parameterizations used for the cross sections, but we do show our prediction for the value of the subtraction function at $Q^2=0$. }
\label{fig:Sinel}
\end{figure}

According to~\eqref{eq:SSigma}, the inelastic part of the subtraction function relevant for the self-energy is determined by the difference between the integrals $\Sigma_1^{\indL}(Q^2)^{p-n}$ and $\Sigma^{\indT}(Q^2)^{p-n}$. The central values of these integrals are shown in \fig~\ref{fig:Sigma1pn}. The narrow band in \fig~\ref{fig:Sinel} indicates the corresponding result for the subtraction function. The width of the band is obtained by evaluating the uncertainties in the contributions arising from the three subintervals, separately for the transverse and longitudinal contributions, and adding the results in quadrature. For better visibility, the vertical axis is stretched with the inverse of the dipole form factor,  $N=(1+Q^2/M_d^2)^2$, $M_d^2=0.71\GeV^2$.  As discussed in \subsec~\ref{subsec:exp}, the region $Q^2<0.5$ contains unphysical fluctuations -- this is why we chop the uncertainty band off there. Note also that, although the calculation returns reasonable results even at $Q^2=2$, it is not reliable there, because it does not account for the contributions by which the AI-parameterization needs to be supplemented in order to agree with experiment at those values of $Q^2$ (see \subsec~\ref{subsec:exp}).

The figure also indicates the value $S_1^\inel(0)^{p-n}= 1.0(2.7)$ obtained from the current experimental result for $\beta_M^{p-n}$, as well as our prediction in~\eqref{eq:Sinel0}. These numbers concern the value of the subtraction function at $Q^2=0$, but are slightly displaced for better visibility. 

\subsection{Comparison with previous work}
\label{sec:comparison}

\bsp
Recently, Walker-Loud, Carlson, and Miller~\cite{WalkerLoud:2012bg} proposed a simple ansatz for the subtraction function. In our notation, their proposal amounts to
\begin{align}
 S_\text{WCM}(q^2)&= -\left(\frac{m_0^2}{m_0^2-q^2}\right)^{2}\frac{\mN\beta_M}{\alphaem}\no
 &+\frac{1}{q^2}\{G_M^2(q^2)-F_D^2(q^2)\}\fs
 \label{eq:ansatz}
\end{align}
The singularity at $q^2=0$ arises from the elastic contribution in~\eqref{eq:Tel}. The corresponding expression for the inelastic part of the subtraction function,\footnote{In~\cite{WalkerLoud:2012bg}, a different terminology is used: there, the first term in~\eqref{eq:ansatzinel} is referred to as the inelastic part of the subtraction function. This is inadequate, as it amounts to counting the polarizability as a purely inelastic quantity. The names given to the various terms are not of importance physically, but when comparing formulae and numerical values, differences in nomenclature must be accounted for.}
\begin{align}
 S_\text{WCM}^\inel(q^2)&=-\left(\frac{m_0^2}{m_0^2-q^2}\right)^{2}\frac{\mN\beta_M}{\alphaem}\no
 &-\frac{4\mN^2\{G_E(q^2)-G_M(q^2)\}^2}{(4\mN^2-q^2)^2}\co
 \label{eq:ansatzinel}
\end{align}
is regular at $q^2=0$ and one readily checks that the ansatz is consistent with the low-energy theorem~\eqref{eq:betainel}. It amounts to an extrapolation of that formula to nonzero values of $q^2$, controlled by the parameter $m_0$. In \fig~\ref{fig:Sinel}, this expression is indicated as a broad band of nearly constant width. 
\esp

\bsp
The ansatz~\eqref{eq:ansatz} for the subtraction function generates a logarithmic divergence in the integral~\eqref{eq:Mgammasub} for the corresponding contribution to the self-energy difference. As discussed in \subsec~\ref{subsec:Lem}, the self-energy difference indeed diverges logarithmically. The divergence is absorbed in the electromagnetic renormalization of $m_u$ and $m_d$, which is of order $e^2m_u$, $e^2m_d$. As pointed out by Erben et al.~\cite{Erben:2014hza}, the logarithmic divergence generated by the ansatz~\eqref{eq:ansatz} is not proportional to the masses of the two lightest quarks and can thus not be absorbed in their renormalization: the particular extrapolation proposed in~\cite{WalkerLoud:2012bg} is not consistent with the short-distance properties of QCD. The variant proposed in~\cite{Erben:2014hza},
\begin{align}
 S_\text{ESTY}^\inel(q^2)&=-\left(\frac{m_1^2-c\,q^2}{m_1^2-q^2}\right)\left(\frac{m_1^2}{m_1^2-q^2}\right)^{2}\frac{\mN\beta_M}{\alphaem}\no
 &-\frac{4\mN^2\{G_E(q^2)-G_M(q^2)\}^2}{(4\mN^2-q^2)^2}\co
 \label{eq:ansatzE}
\end{align} 
repairs this shortcoming, as it disconnects the behaviour at small values of $q^2$ from the asymptotic behaviour. This expression is represented by the central band that gradually shrinks if $Q^2$ increases.
\esp

\bsp
The explicit choice made in~\cite{Erben:2014hza} for the coefficient $c$ implicitly assumes that the contribution from the subtracted dispersion integral, $\mg_\gamma^{\disp}$, stays finite when the cut-off is removed, so that the logarithmic divergence then exclusively arises from the term $\mg_\gamma^S$, which stems from the subtraction function. 
As discussed in detail in~\cite{GL1975}, however, the deep inelastic region also contributes to the coefficient of the logarithmic divergence. The scaling violations do not extinguish this contribution~\cite{Collins}. Hence the choice made for $c$ cannot be taken literally, but it does have the proper quark mass factors, so that the divergence arising from the subtraction function is suppressed. Since the authors cut the integral over the subtraction function off at $\Lambda^2=2\GeV^2$, it barely makes any difference whether $c$ is set equal to zero or taken from~\cite{Erben:2014hza}.
In fact, one of the variants of the model studied in~\cite{Thomas:2014dxa} does correspond to $c=0$.
\esp

\section{Self-energy}
\label{sec:Self-energy}

\subsection{Cottingham formula}
\label{subsec:Lem}

The electromagnetic self-energy of a hadron diverges logarithmically.  To first order in $\alphaem$ the renormalized electromagnetic Lagrangian requires counter terms proportional to the operators $\mathds{1}$, $\qbar q$, and $O_G =G_{\mu\nu}^a G^{a\mu\nu}$:
\begin{align}
 {\cal L}_\text{em}&=-\frac{e^2}{2}\int d^4y \tilde{D}_\Lambda(x-y)Tj^\mu(x)j_\mu(y)+\Delta E\,\mathds{1}\no
 &+\sum_{q=u,d,\ldots}\delta m_q\,\qbar q-\frac{\delta g}{2g^3}O_G\co
 \label{eq:Lem}
\end{align}
where $\tilde{D}_\Lambda(x)$ is the regularized photon propagator in coordinate space.
The counter term proportional to the unit operator does not contribute to the self-energy. The remainder is determined by the renormalization of the quark masses and of the coupling constant $g$ required by the electromagnetic interaction. To leading order, these are given by (see for instance~\cite{GL1982}):
\begin{align}
 \delta m_q&=\frac{3e^2}{16\pi^2}\,\log\frac{\Lambda^2}{\mu^2}\; Q_q^2\,m_q\co\no
\delta g&= -\frac{e^2g^3}{256\pi^4}\,\log\frac{\Lambda^2}{\mu^2}\sum_{q=u,d,\ldots}Q_q^2 \fs
\label{eq:Deltamg}
\end{align}
The form of the regularization used for the photon propagator is irrelevant -- it exclusively affects the value of the running scale $\mu$.

The proton and neutron matrix elements of the operator~\eqref{eq:Lem} lead to a version of the Cottingham formula~\cite{Cottingham} that is valid in QCD: 
\begin{align}
 \mg_\gamma &=\frac{i e^2}{2\mN(2\pi)^4} \int d^4q D_\Lambda(q^2)\{3 q^2 T_1+(2\nu^2+q^2)T_2\}\no
 &+\text{counter terms}\fs
 \label{eq:Mgamma} 
\end{align}
It represents the electromagnetic self-energy in terms of the time-ordered amplitudes $T_1$ and $T_2$ specified in \app~\ref{app:Notation}. 

\subsection{Elastic part of the self-energy}
\label{sec:Elastic part of the self-energy}

Analogously to the electric and magnetic polarizabilities, the self-energy also consists of an elastic and an inelastic part, 
\be
\label{eq:Mgammael}  
\mg_\gamma =  \mg_\gamma^\el + \mg_\gamma^\inel\fs
\ee

\bsp
The contribution from the elastic intermediate states remains finite when the cut-off is removed. It is obtained by replacing $T_1,T_2$ with the elastic parts $T_1^\el,T_2^\el$, which are given explicitly in~\eqref{eq:Tel}, and replacing $D_\Lambda(q^2)$ with the full photon propagator, $D(q^2)=(-q^2-i\epsilon)^{-1}$. With a Wick rotation, the expression can be brought to the form
\begin{align}
 \mg^\el_\gamma &=\frac{\alphaem}{8\pi \mN^3}\int_0^\infty dQ^2 Q^2\{f_1 \,v_1^\el(-Q^2)+f_2 \,v_2^\el(-Q^2)\}\co\no
f_1&= 3\left\{\sqrt{1+\frac{1}{y}}-1\right\}\co\no
f_2&= (1-2y)\sqrt{1+\frac{1}{y}}+2y\co
\label{eq:Mel}
\end{align}
where $v_1^\el(q^2)$ and $v_2^\el(q^2)$ represent the sums of squares of form factors specified in~\eqref{eq:Vel}. The variable $y$ stands for  $y\equiv \nu^2/Q^2$. 
For the elastic contribution, which is concentrated to the line $Q^2=2m\nu$, we have $y=Q^2/4m^2$. In~\cite{GL1975}, the dipole approximation for the Sachs form factors was used, which yields $ \mg_\gamma^\el=0.63\MeV$ for the proton and $-0.13\MeV$ for the neutron, so that the elastic contribution to the self-energy difference amounts to  $(\mg_\gamma^\el)^{p-n} =0.76\MeV$.
\esp

In the meantime, the precision to which the form factors are known has increased significantly. For a thorough review of the experimental information, we refer to~\cite{Formfactors}.  The above estimates of the elastic contributions to the proton and neutron self-energies do receive significant corrections, but the difference between proton and neutron is affected by less than $0.02\MeV$. Compared to the uncertainties in the contributions arising from the deep inelastic region, the departures from the dipole approximation are too small to matter.  
 
\subsection{Inelastic part of the self-energy}
\label{sec:Inelastic part of the self-energy}

\bsp
The inelastic part receives three distinct contributions:
\be
\label{eq:Mgammainel}  
\mg_\gamma^\inel = \mg^S_\gamma+\mg^\disp_\gamma+  \mg^\ct_\gamma\fs
\ee
The term $\mg^S_\gamma$ arises from the subtraction function $S^\inel_1(q^2)$,  $\mg^{\disp}_\gamma$ is given by a dispersion integral over the structure functions, and $\mg^{\ct}_\gamma$ accounts for the fact that the electromagnetic interaction renormalizes the quark masses as well as the coupling constant of QCD. In the above discussion of the polarizabilities, renormalization did not play any role, because these concern the properties of $T_1,T_2$ at low energies. In fact, the inelastic part of the magnetic polarizability exclusively picks up the contribution from the subtraction function specified in~\eqref{eq:betainel}. In the decomposition used in~\eqref{eq:Mgammainel}, we have $\beta^\inel=\beta^S$, $\beta^{\disp} =\beta^{\ct}=0$.  
\esp

\bsp
The term $ \mg^S_\gamma$ is obtained by replacing
$T_1(\nu,q^2)$ in~\eqref{eq:Mgamma} by the subtraction function $S^\inel_1(q^2)$, performing a Wick rotation, and averaging over the directions of the Euclidean momentum. The result reads
\be
\label{eq:Mgammasub} 
\mg_\gamma^S=\frac{3\alphaem}{8\pi \mN}\int_0^{\Lambda^2}dQ^2\,Q^2\,
S^\inel_1(-Q^2)\fs
\ee
This term measures the size of the self-energy arising from the subtraction function (more precisely, the inelastic part thereof -- the remainder is included in $\delta \mg_\gamma^\el$). 
\esp

The second term on the right of~\eqref{eq:Mgammainel} is obtained by replacing the amplitudes $T_1,T_2$ with their inelastic parts $T_1^\inel,T_2^\inel$ and dropping the contribution from the subtraction function in the dispersive representation for $T_1^\inel$. The explicit expression reads
\begin{align}
 \mg_\gamma^{\disp}&=\frac{\alphaem}{2\pi \mN}\int_0^{\Lambda^2}dQ^2 \int_{\nuth}^\infty
d\nu\,\nu\no
&\times\Bigg\{\bigg(f_1-\frac{3}{2y}\bigg) \,V_1(\nu,-Q^2)+f_2 \,V_2(\nu,-Q^2)\Bigg\}\fs
\label{eq:Mgammadisp} 
\end{align}
The term with $3/2y$ makes the difference between the unsubtracted and subtracted dispersion integral over $V_1$: it removes the leading term in the behaviour of $f_1$ when $Q^2$ is held fixed and $\nu$ tends to $\infty$, so that the integral over $\nu$ converges, despite the growth of $V_1$ generated by Reggeon exchange. On the other hand, when $Q^2$ becomes large, the behaviour in the deep inelastic region is relevant. In QCD, the contributions from that region diverge logarithmically if the cut-off is removed. In~\eqref{eq:Mgammadisp}, we have simply cut the integral off at $Q^2=\Lambda^2$ -- this amounts to a regularization of the photon propagator in Euclidean space: $D_\Lambda (-Q^2)=\theta(\Lambda^2-Q^2)/Q^2$.

\bsp In the normalization of the states~\eqref{eq:norm}, the mass shift generated by the counter terms in~\eqref{eq:Lem} is given by
\be
\mg^{\ct}_\gamma=-\sum_{q=u,d,\ldots}\hspace{-0.8em}\frac{\delta m_q}{2\mN}\,\langle p|\qbar q|p\rangle +\frac{\delta g}{4\mN g^3}\langle p|O_G|p\rangle\fs\ee
Neglecting second-order isospin-breaking effects proportional to $e^2(m_u-m_d)$, the proton and neutron matrix elements of operators with isospin zero are the same. Hence the operators $O_G$, $\bar{s}s$, $\bar{c}c$, \ldots drop out in the self-energy difference. Moreover,  isospin symmetry relates the neutron matrix elements of the light quarks to those for the proton, e.g.\ $\langle k|\ubar u| k \rangle^{n}=\langle k|\dbar d| k \rangle^{p}$. Using these properties, the contribution from the electromagnetic renormalization of the quark masses to the self-energy difference can be brought to the form 
\be
\label{eq:DeltaMgammact}  
(\mg^{\ct}_\gamma)^{p-n}  =-\frac{\alphaem}{24\pi \mN}(4m_u-m_d)\log\frac{\Lambda^2}{\mu^2}
\langle p| \ubar u-\,\dbar d |p\rangle\fs 
\ee
The formula shows that the coefficient of the logarithmic divergence is proportional to the masses of the two lightest quarks.
In the chiral limit the divergence disappears altogether: if $u$ and $d$ are taken massless, the self-energy difference approaches a finite limit if the cut-off is removed. In reality,  the contributions from the deep inelastic region do generate a logarithmic divergence, albeit with a small coefficient. An update of the analysis performed in~\cite{GL1975} is needed to account for the scaling violations in the corresponding contributions to the renormalized self-energy difference.  
\esp

\subsection{Numerical evaluation}

In~\cite{WalkerLoud:2012bg}, the contribution from the subtraction function to the self-energy difference is evaluated with $\Lambda^2=2\GeV^2$. According to~\eqref{eq:SSigma}, the inelastic part of the subtraction function is given by the difference between two cross section integrals. The part which involves the transverse cross sections, $\Sigma^{\indT}(Q^2)$, generates a convergent contribution to the formula~\eqref{eq:Mgammasub} for the self-energy. As discussed above, our numerical representation of $\Sigma^{\indT}(Q^2)$ becomes incoherent at values of $Q^2$ below $0.5$, but phase space suppresses that region, so that our estimate, $\mg_\gamma^S(\Sigma^{\indT})\simeq -0.14\MeV$, should be close to the truth (actually, with a coherent representation of the available experimental information, this part could be evaluated rather accurately, even without cutting the integral off). 
The corresponding integral over the longitudinal cross section, $\Sigma_1^{\indL}(Q^2)$, is less sensitive to the shortcomings of the representation we are using (this is why we were able to obtain a rather accurate prediction for the difference between the electric polarizabilities of proton and neutron). Numerically, the contribution from that integral to the self-energy difference is tiny:  $\mg_\gamma^S(\Sigma_1^{\indL})\simeq -0.03\MeV$. In other words: the contributions from the Reggeons do require a subtraction, but taken together with those arising from low energies, the entire contribution from the longitudinal cross section to the subtraction function generates a negligibly small part of the self-energy difference. Together with the number for the contributions from the transverse cross section given above, we obtain  
\be
\label{eq:MgammaS} 
\mg_\gamma^S=-0.17\MeV\fs
\ee
This is to be compared with the number obtained by instead inserting the expression~\eqref{eq:ansatzinel} in formula~\eqref{eq:Mgammasub}. With the central value $\beta_M^{p-n}= -1$ used as input in~\cite{WalkerLoud:2012bg}, we obtain $(\mg_\gamma^S)^\text{WCM}=0.50\MeV$. Keeping all other parts of the calculation in \cite{WalkerLoud:2012bg} as they are, but replacing the ansatz for the subtraction function made there with our prediction, the numerical result for the self-energy difference, $\mg_\gamma^\text{WCM}=1.30\MeV$, is lowered by $0.67\MeV$, so that the central value becomes $\mg_\gamma=0.63\MeV$. Repeating the exercise with the model of~\cite{Erben:2014hza}, i.e.~replacing the expression~\eqref{eq:ansatzinel} by~\eqref{eq:ansatzE}, we instead obtain $(\mg_\gamma^S)^\text{ESTY}=0.20\,\MeV$, so that in this case, the central value $\mg_\gamma^\text{ESTY}=1.04\MeV$ is lowered by $0.37\MeV$, which leads to  $\mg_\gamma=0.67\MeV$. In either case, the early estimate obtained in~\cite{GL1975}, $\mg_\gamma^\text{GL}=0.76(30)$ is confirmed. Comparing their parameterization with recent lattice data on the electromagnetic self-energy difference, the authors of~\cite{WalkerLoud:2012bg} and~\cite{Thomas:2014dxa} obtain results for the difference of the magnetic polarizabilities,
$\beta_M^{p-n}=-0.87(85)$ and $\beta_M^{p-n}=-1.12(40)$, respectively, which is lower than our prediction in~\eqref{eq:betapn}.  The difference reflects the fact that, in \fig~\ref{fig:Sinel}, the bands that correspond to their models run above ours. While these extractions involve a model dependence which is difficult to quantify, there has recently been progress in the direct calculation of the polarizability from the lattice, see~\cite{Chang:2015qxa}.

Note that the momentum dependence of the subtraction function must match the behaviour in the deep inelastic region. Taken by itself, the contribution from the subtraction function is very sensitive to the choice of the cut-off $\Lambda$. As shown in~\cite{GL1975}, the term $\mg_\gamma^{\disp}$ is equally sensitive, but  the sum of the two contributions is nearly independent of $\Lambda$, because the Cottingham formula only contains the very weak logarithmic divergence that is related to the electromagnetic renormalization of the quark masses $m_u$ and $m_d$. As indicated in~\eqref{eq:DeltaMgammact}, the coefficient of the divergence is proportional to these masses and hence very small. Also, it does not come exclusively from the subtraction function. The contributions to $\mg_\gamma^{\disp}$ arising from the deep inelastic region contribute to the coefficient of the logarithmic divergence as well. These were estimated in~\cite{GL1975} on the basis of the data available at the time, which did not show any violations of Bjorken scaling. In the meantime, there has been considerable progress in understanding the properties of the structure functions in the deep inelastic region and there is very clear evidence for scaling violations. For a thorough review of these developments, we refer to~\cite{reviewDIS}. A corresponding update of the results obtained on the basis of the Cottingham formula would be of high interest, also in view of the progress made in calculating electromagnetic self-energies on the lattice, but this goes beyond the scope of the present paper. 

\section{Summary and conclusion}
 \label{sec:Conclusion}
 
 \bsp
 \begin{enumerate}
\item Causality relates the imaginary part of the amplitude for Compton scattering on the nucleon in the forward direction to the cross section of the process $e+N\rightarrow e+\text{anything}$. The relation holds for real photons as well as virtual photons of spacelike momentum, $q^2\leq 0$. The spin-averaged forward scattering amplitude involves two invariants, which we denote by $T_1(\nu,q^2)$ and $T_2(\nu,q^2)$. Their imaginary parts are determined by the transverse and longitudinal cross sections of electron scattering, $\sigma_{\indT}$ and $\sigma_{\indL}$.
\item Regge asymptotics implies that only $T_2(\nu,q^2)$ obeys an unsubtracted fixed-$q^2$ dispersion relation, while the one for $T_1(\nu,q^2)$ requires a subtraction, which represents the value of the amplitude at $\nu=0$: $S_1(q^2)=T_1(0,q^2)$.
The dispersive representation of the spin-averaged forward Compton scattering amplitude thus consists of two parts: an integral over the cross sections $\sigma_{\indT},\sigma_{\indL}$ and an integral over the subtraction function $S_1$. The same also holds for the Cottingham formula, which represents the electromagnetic self-energy of the nucleon in terms of the spin-averaged forward Compton amplitude. 

\item It had been pointed out long ago~\cite{GL1975} that -- unless the Compton amplitude contains a fixed pole at $J=0$ -- the subtraction function is unambiguously determined by the cross sections of electron scattering. We do not know of a proof that the Compton amplitude of QCD is free of fixed poles, but assume that this is the case and refer to this assumption as Reggeon dominance.  As briefly discussed in \subsec~\ref{subsec:Reggeons}, the validity of this hypothesis is questioned in the literature. Indeed, an analysis of the Compton amplitude based on first principles that would determine the behaviour in the Regge region (high energies, low photon virtualities) is not available. If the hypothesis were to fail, this would be most interesting, as it would imply that the known contributions generated by the short-distance singularities and the exchange of Reggeons do not fully account for the high-energy behaviour of QCD.

\item On the basis of  Reggeon dominance, we have derived an explicit representation of the subtraction function in terms of the electron scattering cross sections. The representation requires the asymptotic behaviour of the longitudinal cross section to be known up to contributions that disappear at high energies. 
For the proton Compton amplitude, where Pomeron exchange generates the dominating contribution, the available information does not suffice to reliably evaluate the subtraction function. In the difference between proton and neutron, however, the Pomeron drops out. We have shown that the experimental information available at low photon virtuality does suffice to work out the subtraction function relevant for this difference.

\item In~\cite{WalkerLoud:2012bg}, the electron cross sections $\sigma_{\indT}$, $\sigma_{\indL}$ and the subtraction function $S_1(q^2)$ are instead treated as physically independent quantities. The authors invoke the low-energy theorem that relates the value of the subtraction function at $q^2=0$ to the magnetic polarizability and use experimental information about the latter to pin down the value of the subtraction function at the origin. As direct experimental information about the $q^2$-dependence is not available, the authors construct a model for that. Figure~\ref{fig:Sinel} compares their model with our prediction. As pointed out in~\cite{Erben:2014hza}, the model of~\cite{WalkerLoud:2012bg} is not consistent with the fact that the coefficient of the logarithmic divergence vanishes in the chiral limit. The alternative ansatz for the subtraction function proposed there, which does obey this constraint, is also shown in \fig~\ref{fig:Sinel}. 

\item The authors of~\cite{WalkerLoud:2012bg} use their ansatz for the subtraction function to evaluate the difference between the self-energies of proton and neutron  and obtain $\mg_\gamma^\text{WCM}=1.30(03)(47)\MeV$, significantly higher than the result  obtained in~\cite{GL1975}, $\mg_\gamma^\text{GL}=0.76(30)\MeV$. The difference is blamed on a 'technical oversight' committed in~\cite{GL1975}. This claim is wrong: it suffices to replace their ansatz for the subtraction function with the parameter-free representation used in~\cite{GL1975}, which is spelt out explicitly in~\eqref{eq:SSigma} above. Leaving all other elements of their calculation as they are, the central value for the self-energy difference then drops to $\mg_\gamma=0.63\MeV$, thereby neatly confirming the old result. The same conclusion is reached with the calculation performed in~\cite{Erben:2014hza}.  

\item We emphasize that the present work only concerns low photon virtualities. An update of the analysis carried out in~\cite{GL1975} which accounts for the progress made on the experimental and theoretical sides during the last 40 years -- in particular an evaluation of the contributions from the deep inelastic region which accounts for the violations of Bjorken scaling -- is still missing. 

\item Our representation for the subtraction function also leads to a prediction for the difference between the electric polarizabilities of proton and neutron. The result is given in~\eqref{eq:alphapn}. Using the currently accepted results obtained from the Baldin sum rule, this also determines the difference of the magnetic polarizabilities and, using the comparatively rather precise, known value of the electric polarizability of the proton, we obtain an estimate also for the polarizabilities of the neutron. The result is given in~\eqref{eq:alphanbetan}. 

\item The fact that the results obtained from Reggeon dominance are consistent with experiment and even somewhat more precise amounts to a nontrivial test of the hypothesis that the Compton amplitude is free of fixed poles. Quite apart from the possibility of taking new data at small photon virtuality, an improved representation of the available experimental information on the cross sections would allow us to reduce the uncertainties quite substantially -- in particular, if the deficiencies of the available parameterizations mentioned in \subsec~\ref{subsec:exp} could be removed, the main source of uncertainties in our calculation would immediately disappear.

\item The main problem we are facing with our analysis is that all of the well-established features of electron scattering drop out when taking the difference between proton and neutron: the leading terms of the chiral perturbation series are the same, the contribution from the most prominent resonance, the $\Delta(1232)$, is the same, and the leading asymptotic term due to Pomeron exchange is also the same. Since all of these contributions cancel out, not much is left over. Only a fixed pole could prevent the subtraction function relevant for the difference between proton and neutron from being small. The available data do not exclude the occurrence of a fixed pole, but they indicate that if the phenomenon occurs at all, then the pole must have a rather small residue.  
\end{enumerate}
\esp

\begin{acknowledgements}
\bsp
We would like to thank S.~Alekhin, J.~Bl\"umlein, P.~Bosted, L.~S.~Brown, I.~Caprini, E.~Christy, M.~D\"oring, H.~Grie\ss hammer, D.~Haidt, M.~Hilt, G.~Ingelman, B.~Kubis, F.J.~Llanes-Estrada, J.~McGovern, U.-G.~Mei\ss ner, C.~Merino, O.~Nachtmann, D.~R.~Phillips, D.~R\"onchen, S.~Scherer, A.~Thomas, L.~Tiator, and A.~Tkabladze for helpful discussions and correspondence. In particular, we  thank P.~Bosted for making the pertinent Fortran code available to us. J.~G.\ thanks the Helmholtz-Institut f\"ur Strahlen-und Kernphysik at Rheinische Friedrich-Wilhelms-Universit\"at Bonn for hospitality -- part of this work was performed there.   
Financial support by BMBF ARCHES, the Helmholtz Alliance HA216/EMMI, the Swiss National Science Foundation, 
the Deutsche Forschungsgemeinschaft (CRC 16, ``Subnuclear Structure of Matter'' and CRC 110, ``Symmetries and the Emergence of Structure in QCD''),
the Volkswagenstiftung under contract no.\ 86260, and the DOE (Grant No.\ DE-FG02-00ER41132) is gratefully acknowledged.
\esp
\end{acknowledgements}

\appendix

\section{Notation}
\label{app:Notation} 

\bsp
The structure functions are related to the Fourier transform of the spin-averaged matrix element of the current commutator,
\be  
V^{\mu\nu}(p,q)=\frac{1}{4\pi}\int d^4x e^{i q\cdot x}\langle p|[j^\mu(x),j^\nu(0)]|p\rangle\fs
\ee
The states are normalized with 
\be\label{eq:norm}
\langle p',s'|p,s\rangle= 2\hspace{0.02cm}p^0 (2\pi)^3 \delta^3({\bf p}^{\,\prime}-{\bf p})\delta_{s's}
\ee
and $\langle p|O|p\rangle$ stands for $\frac{1}{2}\sum_s \langle p,s|O|p,s\rangle$.
For the Fourier transform of the time-ordered matrix element, $\langle p|Tj^\mu(x)j^\nu(y)|p\rangle$, we use the normalization
\be
\label{eq:normTi}  
T^{\mu\nu}(p,q)=\frac{i}{2}\int d^4x e^{i q\cdot x}\langle p|Tj^\mu(x)j^\nu(0)|p\rangle\fs
\ee
Since Lorentz invariance, current conservation, and parity only allow two independent tensors of this type,
\begin{align}
 K_1^{\mu\nu}&= q^\mu q^\nu-g^{\mu\nu}q^2\co\no
 K_2^{\mu\nu}&=  \frac{1}{\mN^2}\Big\{(p^\mu q^\nu+p^\nu q^\mu)p\cdot q\no
 &\qquad-g^{\mu\nu}(p\cdot q)^2-p^\mu p^\nu q^2\Big\}\co
 \label{eq:K}
\end{align}
 these matrix elements contain two invariants each, which only depend on the two variables $\nu\equiv p\cdot q/\mN$ and $q^2$ ($\mN$ is the mass of the nucleon). We denote the invariants by $V_1(\nu,q^2), V_2(\nu,q^2)$ and $T_1(\nu,q^2), T_2(\nu,q^2)$, respectively:
 \begin{align}
  V^{\mu\nu}(p,q)&= V_1(\nu,q^2)K_1^{\mu\nu} +V_2(\nu,q^2)K_2^{\mu\nu} \co\no
 T^{\mu\nu}(p,q)&= T_1(\nu,q^2)K_1^{\mu\nu} +T_2(\nu,q^2)K_2^{\mu\nu} \fs
  \label{eq:defVandT} 
 \end{align}
 In contrast to the standard structure functions $F_1$, $F_2$ the invariants $V_1$, $V_2$ are free of kinematic singularities or zeros. 
 The two sets are related by 
\be
\label{eq:F1F2}
V_1\equiv\frac{-2x F_1+F_2}{4\mN x^2\nu}\co\quad 
V_2\equiv\frac{ F_2}{2x\nu^2 }\co
\ee
with $x\equiv Q^2/2\mN\nu$.
\esp

The notation for the longitudinal structure function $F_{\indL}$ is not universal. The convention used in the mini-review on the structure functions in The Review of Particle Physics~\cite{StructureFunctions} reads
\be
\label{FL}
F_{\indL}=F_2-2x F_1\fs
\ee 

The structure functions $V_1(\nu,q^2)$ and $V_2(\nu,q^2)$ represent linear combinations of the transverse and longitudinal cross sections $\sigma_{\indT}$ and $\sigma_{\indL}$:
\begin{align}
 V_1&= N_1(-Q^2 \sigma_{\indT}+\nu^2 \sigma_{\indL})/Q^2\co\quad
 V_2=N_1(\sigma_{\indT}+\sigma_{\indL})\co\no
N_1&\equiv\frac{1}{8\pi^2\alphaem}\frac{2\mN\nu-Q^2}{\nu^2+Q^2}\fs
\label{eq:Vsigma}
\end{align}
The value of the structure function $V_2(\nu,0)$ also determines the total cross section for photoproduction, $\sigma_\text{tot}=\lim_{Q^2\rightarrow 0}\sigma_{\indT}$:
\be 
\sigma_\text{tot}(\nu)=4\pi^2 \alphaem\frac{\nu}{\mN} V_2(\nu,0)\fs
\ee

\bsp
For the one-particle matrix elements of the current, we use the notation 
\begin{align}
 \langle p_1,s_1|j^\mu |p_2,s_2\rangle &= \bar{u}(p_1,s_1)\Gamma^\mu(q)u(p_2,s_2)\co\no
\Gamma^\mu(q)&= F_D(q^2)\gamma^\mu+  F_P(q^2)i \sigma^{\mu\nu}\frac{q_\nu}{2\mN}\co
 \label{eq:FDFP} 
\end{align}
where $q=p_1-p_2$. The nucleon
spinors are normalized with $\ubar(p,s') u(p,s) = 2 \mN\delta_{s's}$.
The functions $F_D(t)$ and $F_P(t)$ are referred to as Dirac and Pauli form factors, respectively. Whenever convenient, we replace these by the Sachs form factors, which are defined by 
\be
\label{eq:Dirac}
G_E(t)=F_D(t)+ \frac{t}{4\mN^2}F_P(t)\co\quad
G_M(t)=F_D(t)+ F_P(t)\fs
\ee 
In dipole approximation, the form factors are parameterized with
\begin{align}
 G_E^p(t)&= G_d(t)\co\quad G_M^p(t)=(1+\kappa^p)G_d(t)\co\no
G_E^n(t)&=\frac{t}{4\mN^2}\kappa^nG_d(t)
\co\quad G_M^n(t)=\kappa^nG_d(t)\co\no
G_d(t)&=\frac{1}{\left(1-t/M_d^2\right)^2}\co\quad
M_d^2\simeq 0.71 \GeV^2\co
\label{eq:Dipole}
\end{align} 
where $\kappa=F_P(0)$ stands for the anomalous magnetic moment.
\esp

\section{Compton scattering}
\label{app:LET}

\bsp
Virtual  Compton scattering in the non-forward direction provides the bridge between the two processes discussed in \subsec~\ref{subsec:Low energy theorem}: scattering of real photons at nonzero scattering angle and scattering of virtual photons in the forward direction. Compton scattering has been thoroughly explored in the literature, for the case where both of the two photons are on the mass shell (real Compton scattering, RCS) as well as when one of them (VCS) or both (VVCS) are off-shell~\cite{Tarrach,Gorchtein:2008aa,Birse:2012eb,Drechsel,Hill:2011wy,
Berg,Saito,Bernabeu:1976jq,Petrunkin:1981me,Lvov:1993fp,Guichon:1995pu,Fearing:1996gs,Hemmert:1996gr,Drechsel:1996ag,Hemmert:1997tj,Drechsel:1998zm,Pachucki:1999zza,Drechsel:2002ar,Wissmann:2004ny,Schumacher:2005an,Gorchtein:2009wz,Holstein:2013kia}. We normalize the amplitude with
\begin{align} 
\hat{T}^{\mu\nu}(p_f, s_f, q_f|p_i,s_i, q_i)
&=\frac{i}{2}\int d^4x e^{i\, q_f\cdot x} \no
&\times\langle p_f,s_f|Tj^\mu(x)j^\nu(0)|p_i,s_i\rangle\co
\end{align}
and use matrix notation, collecting the different spin orientations in the $2\times 2$ matrix $\boldsymbol{T}^{\mu\nu}(p_f,q_f|p_i,q_i)$. 
The spin average is given by the trace of this matrix,
\be
\label{eq:spin average}  
T^{\mu\nu}(p_f,q_f|p_i,q_i)=\frac{1}{2}\mbox{tr}\left\{\boldsymbol{T}^{\mu\nu}(p_f,q_f|p_i,q_i) \right\}\fs
\ee
\esp

\subsection{Lorentz invariance}
\label{app:Lorentz}

\bsp
The spin average is not independent of the Lorentz frame used. To see why this is so, consider a Lorentz transformation:
\be
\label{eq:Uj} 
U(\Lambda)j^\mu(x)U^{-1}(\Lambda)=(\Lambda^{-1})^\mu_{\,\alpha}\,j^\alpha(\Lambda  x)\fs
\ee
We denote the pure Lorentz transformation (boost) that takes a particle at rest into one of four-momentum $p$ by $B_p$ and work in the basis where the state $|p,s\rangle$ is obtained from the corresponding state at rest, $|\hat{p},s\rangle$, by application of the relevant boost: $|p,s\rangle=U(B_p)|\hat{p},s\rangle$. Lorentz transformations not only change the momentum, but also subject the spin direction to a rotation, referred to as {\it Wigner rotation}: 
\be
\label{eq:Up} 
U(\Lambda)|p,s\rangle=\sum_{s'}|\Lambda p,s'\rangle \hat{W}_{s's}(\Lambda, p)\fs
\ee
The Wigner rotation arises because the boost $B_{\Lambda p}$ differs from $\Lambda B_p$ by a rotation, which we denote by $W(\Lambda,p)$: 
\be 
\label{eq:Wigner}
\Lambda B_p=B_{\Lambda p} W(\Lambda,p)\fs
\ee 
The matrix $\hat{W}_{s's}(\Lambda,p)$ in~\eqref{eq:Up} is the spin $\frac{1}{2}$ representation of $W(\Lambda,p)$. 
If $\Lambda$ is a pure rotation, we have $W(R,p)=R$. Also, since the product of two boosts in the same direction is again a boost in that direction, a pure Lorentz transformation in the direction of $\boldsymbol{p}$ does not generate a Wigner rotation. 
\esp

Lorentz invariance implies the transformation law
\begin{align}
\label{eq:trafoW}  
&\Lambda^\mu_{\,\alpha}\Lambda^\nu_{\,\beta}\boldsymbol{T}^{\alpha\beta}(p_f,q_f|p_i,q_i)\no
&=\boldsymbol{W}^{\dagger}(\Lambda,p_f)\boldsymbol{T}^{\mu\nu}(\Lambda p_f,\Lambda q_f|\Lambda p_i,\Lambda q_i) \boldsymbol{W}(\Lambda, p_i)\fs
\end{align}
In the trace, the Wigner rotations only drop out for those Lorentz transformations for which $W(\Lambda,p_i)=W(\Lambda,p_f)$. In general, this condition is violated. Hence knowledge of the spin average in one particular frame of reference does not in general suffice to determine the spin average in a different frame: the transformation law~\eqref{eq:trafoW} involves the entire matrix  $\hat{T}^{\mu\nu}$, including the spin-flip components of the amplitude.

In the Breit frame, $\boldsymbol{p}_f+\boldsymbol{p}_i =0$,  the momenta of the initial and final states point in opposite directions. The boost which takes the Breit frame into the Lab frame, where $\boldsymbol{p}_i=0$, is a pure Lorentz transformation in that direction. Hence the change of frame does not generate a Wigner rotation. Accordingly, the spin average in the Lab is determined by the spin average in the Breit frame. 

For the Lorentz transformation that takes the Breit frame into the centre-of-mass system, however, this is not the case: the Wigner rotation generated by this transformation for the initial state differs from the one relevant for the final state, $W(\Lambda, p_i)\neq W(\Lambda,p_f)$. Hence knowledge of the spin average in the Breit frame does not suffice to evaluate the spin average in the centre-of-mass system or vice versa.  

\subsection{Crossing symmetry, parity, and time reversal}
\label{app:CPT}

\bsp
The symmetry of the time-ordered product, $Tj^\mu(x)j^\nu(y)=Tj^\nu(y)j^\mu(x)$, implies invariance under crossing of the photons:\footnote{If the time-ordered product of the currents is replaced by the retarded current commutator, crossing symmetry instead relates the amplitude to its complex conjugate~\cite{Tarrach}.} 
\be
\label{eq:crossingT}
\boldsymbol{T}^{\mu\nu}(p_f,q_f|p_i,q_i)=\boldsymbol{T}^{\nu\mu}(p_f,-q_i|p_i,-q_f)\fs
\ee
Invariance under space reflections amounts to:
\be
\label{eq:parityT}
\boldsymbol{T}^{\mu\nu}(p_f,q_f|p_i,q_i)=\pi^{\mu}_{\,\alpha}\pi^{\nu}_{\,\beta}\,  
\boldsymbol{T}^{\alpha\beta}(\pi p_f,\pi q_f|\pi p_i,\pi q_i) \co
\ee
where $\pi=\mbox{diag}(1,-1,-1,-1)$ inverts the sign of the space components  but leaves the time components alone. 
Time reversal not only inverts the momentum and spin directions, but in addition interchanges the initial and final states. Moreover, the amplitudes are mapped into their complex conjugate. Exploiting the fact that the Hermitian conjugate of the operator $j^\mu(x)j^\nu(y)$ is given by $j^\nu(y)j^\mu(x)$, time reversal invariance can be brought to the form
\begin{align}
\label{eq:timereversalT}
&\boldsymbol{T}^{\mu\nu}(p_f,q_f|p_i,q_i)\no
&=\pi^{\nu}_{\,\alpha}\pi^{\mu}_{\,\beta}\, \epsilon\,\boldsymbol{T}^{\alpha\beta}(\pi p_i,\pi q_i|\pi p_f,\pi q_f)^{\indT}\epsilon^{-1}\fs
\end{align}
The superscript $T$ indicates that the transposed matrix is relevant. The matrix $\epsilon = i \sigma_2$ flips the spin in the initial and final states. For the Pauli matrices, we have $\epsilon\,\boldsymbol{\sigma}^{\indT}\epsilon^{-1}=-\boldsymbol{\sigma}$. 
The above relations lead to the following symmetry property of  $\boldsymbol{T}^{\mu\nu}$:
\be
\label{eq:Symmetry}
\boldsymbol{T}^{\mu\nu}(p_f,q_f|p_i,q_i)=\epsilon\,\boldsymbol{T}^{\mu\nu}(p_i,-q_f|p_f,-q_i)^{\indT}\epsilon^{-1}\fs
\ee
It implies that, in the decomposition  $\boldsymbol{T}^{\mu\nu}=T^{\mu\nu}\boldsymbol{1}+\sum_{i} T_i^{\mu\nu}\boldsymbol{\sigma}^i$ of the amplitude in the basis spanned by $\boldsymbol{1},\boldsymbol{\sigma}_1,\boldsymbol{\sigma}_2,\boldsymbol{\sigma}_3$, the spin-independent part, $T^{\mu\nu}$, is even under the operation $p_i\leftrightarrow p_f$, $q_i\rightarrow -q_i$, $q_f\rightarrow -q_f$, while the spin-dependent part, $T_i^{\mu\nu}$, is odd. The mapping interchanges the Mandelstam variables $s=(p_i+q_i)^2$ and $u=(p_i-q_f)^2$, but $t=(p_f-p_i)^2$ as well as the photon virtualities $q_i^2,q_f^2$ stay put. While the Breit frame is invariant under this operation, neither the Lab frame nor the centre-of-mass system have that property.
\esp

\subsection{Low-energy expansion}
\label{app:Derivation}

The elastic intermediate states generate poles in $\boldsymbol{T}^{\mu\nu}$. In the Mandelstam variables the poles are located at $s=\mN^2$ and $u=\mN^2$. We refer to these contributions as Born terms and denote them by $\boldsymbol{T}^{\mu\nu}_B$,
\be
\boldsymbol{T}^{\mu\nu}=\boldsymbol{T}^{\mu\nu}_B+\boldsymbol{\bar{T}}^{\mu\nu}\fs
\ee
The decomposition is not  unique~\cite{Scherer:1996ux,Fearing:1996gs,Birse:2012eb}. The essential property of the Born terms is that they account for the elastic singularities. This ensures that the remainder, $\boldsymbol{\bar{T}}^{\mu\nu},$ is regular at $q_i=q_f=0$ and can thus be expanded in a Taylor series in powers of the photon  momenta and energies, which the low-energy expansion treats as small. 
The construction described below leads to Born terms that are conserved, so that this also holds for the remainder:\footnote{In general, it is 
not a trivial matter to impose current conservation on the Born terms. If one for instance evaluates the one-particle singularities in the space components $\boldsymbol{T}^{ab}_B$ and determines the remaining components of $\boldsymbol{T}^{\mu\nu}_B$ by solving the constraints $q_{\mu}^f\,\boldsymbol{T}^{\mu\nu}_B=q_{\nu}^i\,\boldsymbol{T}^{\mu\nu}_B=0$, one in general arrives at a representation for the Born terms that contains kinematic singularities. The presence of kinematic singularities in the Born terms complicates the analysis because the remainder $\boldsymbol{\bar{T}}^{\mu\nu}$ is then not regular at $q_i=q_f=0$. Compton scattering on the pion illustrates the problem: the one-particle singularities do not generate a term proportional to $g^{\mu\nu}$, but unless a regular term of this type is allowed for, the representation of the Born terms can be consistent with current conservation only if it contains kinematic singularities.} 
\be
\label{eq:Tbarconserved} 
q_{\mu}^f\,\boldsymbol{\bar{T}}^{\mu\nu}=q_{\nu}^i\,\boldsymbol{\bar{T}}^{\mu\nu}=0\fs
\ee

Keeping $\boldsymbol{P}\equiv\frac{1}{2}(\boldsymbol{p}_i+\boldsymbol{p}_f)$ fixed, momentum  conservation determines the initial and final nucleon momenta in terms of the photon variables $\boldsymbol{q}_i,\boldsymbol{q}_f$. However, unless $\boldsymbol{P}$ vanishes, energy conservation leads to a nonlinear constraint on the photon energies, so that it is not consistent to treat all of these as quantities of $O(q)$. The problem disappears for $\boldsymbol{P}=0$, i.e.\ in the Breit frame: energy conservation then implies $\omega_i=\omega_f=\omega$, so that the kinematics is unambiguously determined by the independent variables $\omega, \boldsymbol{q}_i, \boldsymbol{q}_f$, which all count as quantities of $O(q)$. This is why the Breit frame is the preferred frame of reference for the low-energy expansion. For real Compton scattering, a transparent discussion of this issue is given in~\cite{Babusci}.

In the Breit frame, the  Taylor series in powers of the variables $\omega, \boldsymbol{q}_i, \boldsymbol{q}_f$  takes the form:
\be
\boldsymbol{\bar{T}}^{\mu\nu}=\boldsymbol{\bar{T}}^{\mu\nu}_0+\boldsymbol{\bar{T}}^{\mu\nu}_1+\boldsymbol{\bar{T}}^{\mu\nu}_2+\ldots\quad
\text{with}\quad \boldsymbol{\bar{T}}^{\mu\nu}_n=O(q^n)\fs
\ee 
The symmetry~\eqref{eq:Symmetry} implies that the even terms of the series $\boldsymbol{\bar{T}}^{\mu\nu}_0,\boldsymbol{\bar{T}}^{\mu\nu}_2,\ldots$ are proportional to the unit matrix in spin space, while the odd terms $\boldsymbol{\bar{T}}^{\mu\nu}_1,\boldsymbol{\bar{T}}^{\mu\nu}_3,\ldots$ exclusively contain spin-dependent terms and do not contribute to the spin average.

We now turn to the consequences of Lorentz invariance for the low-energy expansion. If the photon energies and momenta are small of $O(q)$, the momentum transferred to the nucleon is small as well. We thus only need to consider Lorentz frames where the nucleon momenta are also small of $O(q)$. The standard choices (Laboratory, centre-of-mass system, Breit frame) all belong to this category. If a Lorentz transformation  $\Lambda$ is to connect two such frames, then the relative velocity must be small, so that the standard decomposition into a boost and a rotation, $\Lambda=BR$ only involves a small boost. Hence it suffices to analyze the transformation properties under rotations and under small boosts. 

The behaviour under rotations is trivial, because the corresponding Wigner rotations in the initial and final states are identical, $W(R,p_i)=W(R,p_f)=R$. Hence they leave the spin-independent part of the amplitude alone and transform the matrices $\boldsymbol{\sigma}=
\{\sigma_1,\sigma_2,\sigma_3\}$ occurring in the spin-dependent part like a vector. 

\bsp
To analyze the properties of small boosts, the $SL(2,C)$ representation of the Lorentz group is more convenient than the one acting on the coordinates and momenta. Consider the matrix $B=\exp(\frac{1}{2}\boldsymbol{v}\cdot\boldsymbol{\sigma}) \in SL(2,C)$, which represents a pure Lorentz transformation with a small velocity $\boldsymbol{v}=O(q)$. The product of two such boosts is given by $B B'=\boldsymbol{1}+\frac{1}{2}\boldsymbol{w}\cdot\boldsymbol{\sigma} +\frac{1}{8}\boldsymbol{w}^2+\frac{i}{4}(\boldsymbol{v}\times \boldsymbol{v}')\cdot\boldsymbol{\sigma}+O(q^3)$, with  $\boldsymbol{w}=\boldsymbol{v}+\boldsymbol{v}'$. At $O(q)$, this is a pure Lorentz transformation,\footnote{The fact that the velocity addition is modified only shows up at $O(q^3)$.} but at  $O(q^2)$, the product in addition contains a small Wigner rotation: $W=\mbox{\bf 1}+\frac{i}{4}(\boldsymbol{v}\times \boldsymbol{v}')\cdot\boldsymbol{\sigma}+O(q^4)$. 
\esp

The essential point here is that, for the pure Lorentz transformation needed to remove the relative velocity of the two systems, the Wigner rotations generated in the initial and final states are at most of order $q^2$. Accordingly, the Lorentz invariance condition~\eqref{eq:trafoW} implies that, if $\Lambda$ is a boost that only generates velocities of $O(q)$, the individual terms of the low-energy expansion transform like ordinary tensors -- up to higher order corrections:  
\begin{align}
\label{eq:trafoE} 
B^\mu_{\,\alpha}B^\nu_{\,\beta}\boldsymbol{\bar{T}}_n^{\alpha\beta}(p_f,q_f|p_i,q_i)&=\boldsymbol{\bar{T}}_n^{\mu\nu}(B p_f,B q_f|B p_i,B q_i)\no &+O(q^{n+2})\fs
\end{align}
The  corrections only matter if the expansion is taken beyond next-to-leading order. In particular,  the leading even term of the series is proportional to the unit matrix in spin space even if the reference system is not identified with the Breit frame.  

\subsection{Spin average}
\label{app:Spin}

\bsp
The Breit frame also offers a convenient decomposition of the amplitude into independent tensors. In the following, we explicitly construct the decomposition for the spin average, which we denote by $T^{\mu\nu}$.  Current conservation ($q_{\mu}^f\,T^{\mu\nu}=q_{\nu}^i\,T^{\mu\nu}=0$) implies that the amplitude is uniquely determined by its space components. Furthermore, rotation invariance ensures that the spin average involves five independent amplitudes ($a,b = 1,2,3$): 
\be
\label{eq:t_n} 
T^{ab}=\delta^{ab}\,I_1 +q^a_i q^b_f \,I_2+q^a_i q^b_i \,I_3+q^a_fq^b_f \,I_4+q^a_f q^b_i \,I_5\fs
\ee
Since this decomposition exclusively makes use of rotation invariance, the coefficients $I_n$ are free of kinematic singularities. They depend on the rotation-invariant quantities $\omega, |\boldsymbol{q}_i|, |\boldsymbol{q}_f|$, and $ \boldsymbol{q}_i\cdot \boldsymbol{q}_f$, which can be  expressed in terms of the Mandelstam variables and the photon virtualities. In view of $s+t+u=2\mN^2 +q_i^2+q_f^2$, only four of these are independent, for instance: $s,u,q_i^2,q_f^2$. 
\esp

\bsp
The trace of the relation~\eqref{eq:Symmetry} implies that the invariants are symmetric under the interchange of $s$ and $u$, 
\be
\label{eq:su}
I_n(s,u,q_i^2,q_f^2)=I_n(u,s,q_i^2,q_f^2)\co\quad n=1,\ldots,5\co
\ee
and the crossing symmetry relation~\eqref{eq:crossingT} then shows that they are symmetric under $q_i^2\leftrightarrow q_f^2$ as well, except that $I_3$ and $I_4$ are interchanged,
\begin{align}
I_n(s,u,q_i^2,q_f^2)&=I_n(s,u,q_f^2,q_i^2)\co\quad  n=1,2,5\co\no
I_3(s,u,q_i^2,q_f^2)&=I_4(s,u,q_f^2,q_i^2)\fs
\label{eq:qiqf}
\end{align} 
The last relation implies that if the photon virtualities are the same -- in particular for real Compton scattering -- there are only four independent amplitudes~\cite{Tarrach}.
\esp

\bsp
We add a remark concerning the spin average for the case where the scattering amplitude is written in the form  
\begin{align}
 &\hat{T}^{\mu\nu}(p_f,s_f,q_f|p_i,s_i,q_i)\no
 &=\bar{u}(p_f,s_f)M^{\mu\nu}(p_f,q_f|p_i,q_i)u(p_i,s_i)\fs
 \end{align}
If the momenta $\boldsymbol{p}_i$ and $\boldsymbol{p}_f$ are parallel, the sum over the spins can be represented as a product of projectors:
\begin{align}
 \sum_s u(p_i,s)\otimes\bar{u}(p_f,s)&=\frac{1}{K}\,
 (\slashed{p}_i+\mN)\cdot(\slashed{p}_f+\mN)\co\no
 K&=2\mN\sqrt{1-t/4\mN^2}\co
 \label{eq:spinsum}
\end{align}
so that spin average can be represented as 
 \begin{align}
   &T^{\mu\nu}(p_f,q_f|p_i,q_i)\no
   &=\frac{1}{2K}\mbox{tr}\left\{(\slashed{p}_f+\mN)\cdot M^{\mu\nu}(p_f,q_f|p_i,q_i)\cdot
(\slashed{p}_i+\mN)\right\}\fs
\label{eq:TM}
 \end{align}
This formula might suggest that a frame-independent definition of the spin average does exist. As already noted by Tarrach~\cite{Tarrach}, this is not the case, however: the relation~\eqref{eq:TM} is correct only in those Lorentz frames where $\boldsymbol{p}_i$ and $\boldsymbol{p}_f$ are parallel. In particular, it does not hold in the centre-of-mass frame, where the right hand side of~\eqref{eq:TM} does not represent the spin average performed on the left hand side.
\esp

\subsection{Born terms}
\label{app:Born_terms}

\bsp
The Born terms contain poles along the lines $s=\mN^2$ and $u=\mN^2$. The residue of the pole in the $s$-channel involves a sum over the one-particle matrix elements of the current: 
\be
\sum_s\bar{u}(p_f,s_f)\Gamma^\mu(-q_f)u(p_n,s)\bar{u}(p_n,s) \Gamma^\nu(q_i)u (p_i,s_i)\fs
\ee
At the poles, the momentum of the intermediate state is on the mass-shell, $p_n^2=\mN^2$, and the sum over the spin directions is given by 
\be
\sum_s u(p_n,s)\,\bar{u}(p_n,s)= \slashed{p}_n+\mN\fs
\ee
We specify the residues off the mass-shell with analytic continuation, simply replacing $p_n$ with the total momentum, $P=p_i+q_i=p_f+q_f$, also for $P^2\neq \mN^2$: 
\begin{align}
 \hat{T}^{\mu\nu}_B(p_f,s_f,q_f|p_i,s_i,q_i)&= \frac{\hat{B}^{\mu\nu}(p_f,s_f,q_f|p_i,q_i)}{\mN^2-s}\no
 &+\frac{\hat{B}^{\nu\mu}(p_f,s_f,-q_i|p_i,s_i,-q_f)}{\mN^2-u}\co\no
\hat{B}^{\mu\nu}(p_f,s_f,q_f|p_i,s_i,q_i)& = \frac{1}{2}\bar{u}(p_f,s_f)\Gamma^\mu(-q_f)\no
&\times(\slashed{P}+\mN)\Gamma^\nu(q_i) u(p_i,s_i)\fs
\end{align}
The two terms represent the two tree graphs obtained with the standard Feynman rules, except that the photon-nucleon vertices are equipped with form factors according to~\eqref{eq:FDFP}. Taken separately, the two terms do not obey current conservation, but taken together they do. 
\esp

\bsp
In the Breit frame, the spin average can be evaluated with~\eqref{eq:spinsum}. Comparing the resulting expression for the space components with~\eqref{eq:t_n}, we obtain the following explicit representation for the Born terms:
\begin{align}
 I_n^B&= \frac{\sqrt{1-t/4\mN^2}}{(\mN^2-s)(\mN^2-u)}B_n\co\quad n=1,\ldots\,, 5\co\no
B_1&= 4\mN^2\omega^2F_D^i F_D^f +  w_1(F_D^i F_P^f + F_D^f F_P^i+  F_P^i F_P^f)\co\no
B_2& =  w_2 (F_D^i F_P^f + F_D^f F_P^i+ F_P^i F_P^f)\co\no
B_3&= \omega^2F_P^f(F_D^i+ F_P^i)\co\no
B_4&=\omega^2 F_P^i(F_D^f+  F_P^f)\co\no
B_5&= 0\co\no
w_1&\equiv  \frac{1}{4}(q_i^2+q_f^2)^2- \frac{1}{4}t^2-t\, w_2\co\no
w_2&\equiv \frac{1}{2}(q_i^2+q_f^2)- \frac{1}{2}t-\omega^2\fs
\label{eq:tB} 
\end{align}
The symbol $F_D^i$ stands for $F_D(q_i^2)$ and $F_D^f, F_P^i, F_P^f$ are defined analogously. In the Breit frame, the variable $\omega$ represents the energy of the photons. Expressed in terms of the Mandelstam variables, we have 
\be
\omega=\frac{s-u}{2\sqrt{4\mN^2-t}}\fs
\ee
These expressions of course  satisfy the relations~\eqref{eq:su} and~\eqref{eq:qiqf}. In the case of equal photon virtualities, Born terms for the spin-averaged amplitude are also provided in~\cite{Birse:2012eb}. The representation specified in~(3) and (4) of that work differs from ours in the overall normalization of the amplitude: basically, it amounts to replacing the factor $K$ in~\eqref{eq:spinsum} by $2\mN$.  
\esp

\subsection{Leading low-energy constants}
\label{app:Leading}

Since $\omega, \boldsymbol{q}_i, \boldsymbol{q}_f$ are independent variables and $\boldsymbol{\bar{T}}^{\mu\nu}_0$ is independent thereof, the constraint $q_\mu^f\boldsymbol{\bar{T}}^{\mu\nu}_0=0$ immediately implies $\boldsymbol{\bar{T}}^{\mu\nu}_0=0$. The term $\boldsymbol{\bar{T}}^{\mu\nu}_1$ is a linear combination of the variables $\omega, \boldsymbol{q}_i, \boldsymbol{q}_f$, but this property is inconsistent with current conservation: since invariance under space reflections requires the components $\boldsymbol{\bar{T}}_1^{00}$ and $\boldsymbol{\bar{T}}_1^{ab} $ to be even under a reversal of the photon momenta, they must be independent thereof. The constraint $\omega^2 \boldsymbol{\bar{T}}_1^{00}=q_f^aq_i^b \boldsymbol{\bar{T}}_1^{ab}$ can then only be obeyed if $\boldsymbol{\bar{T}}_1^{00}$ and  $\boldsymbol{\bar{T}}_1^{ab}$ both vanish, but this is compatible with current conservation only if the remaining components, $\boldsymbol{\bar{T}}_1^{0a},\boldsymbol{\bar{T}}_1^{a0}$, also vanish. 

Accordingly, the low-energy expansion of $\boldsymbol{\bar{T}}^{\mu\nu}$ only starts at $O(q^2)$. Indeed, the following calculation shows that there are exactly two independent conserved tensors of that order.  As noted above, the contribution of $O(q^2)$ is spin independent, $\boldsymbol{\bar{T}}_2^{\mu\nu} =T_2^{\mu\nu}\,\boldsymbol{1}$ and rotation invariance requires the space components 
$\bar{T}^{ab}_2$ to be of the form~\eqref{eq:t_n} with $\bar{I}_1=c_1\,\omega^2+c_2 \,\boldsymbol{q}_f\cdot \boldsymbol{q}_i+c_3\,\boldsymbol{q}_i^{2}+c_4\, \boldsymbol{q}_f^{2}$, while the other coefficients are constants. For $\omega=0$, current conservation requires $q_f^a \bar{T}^{ab}_2=q_i^b\bar{T}^{ab}_2=0$. This implies $\bar{I}_2=-c_2$, $\bar{I}_3=\bar{I}_4=\bar{I}_5=c_3=c_4=0$. For the space components, the general solution of the conditions imposed by rotation invariance and current conservation at $O(q^2)$ thus reads
\be
\label{eq:LEC} 
\bar{T}^{ab}_2=c_1 \,\delta^{ab} \omega^2 + c_2\,\{\delta^{ab}\boldsymbol{q}_f\cdot\boldsymbol{q}_i-q^a_iq^b_f \} \fs
\ee
Current conservation then fixes the remaining components in terms of the same two constants:
\be
\label{eq:Tbar0}
\bar{T}_2^{a0}=c_1\, \omega\, q_i^a\co\quad \bar{T}_2^{0a}=c_1\,\omega\, q_f^a\co\quad \bar{T}_2^{00}=c_1\,\boldsymbol{q}_f\cdot \boldsymbol{q}_i\fs
\ee
As observed already by Klein~\cite{Klein:1955zz}, there are only two conserved tensors of polynomial form at $O(q^2)$.
  
\subsection{Real Compton scattering}
\label{app:real_Compton}

\bsp
For real photons, $q_i^2=q_f^2=0$, the projection onto the polarization vectors ($\epsilon^0=0$, $\boldsymbol{\epsilon}\cdot\boldsymbol{q}=0$) annihilates the time components as well as the contributions from $I_3,I_4,I_5$.  At low frequencies, the scattering amplitude is fully determined by the charge of the particle~\cite{Thirring}, which we express in units of the proton charge, $Q_e\equiv F_D(0)$. At first order in the expansion in powers of the photon frequency $\omega$,  the anomalous magnetic moment $\kappa=F_P(0)$ also shows up~\cite{Low,Gell-MannGoldberger}, and at $O(\omega^2)$, further contributions, characterized by the two low-energy constants in~\eqref{eq:LEC}, manifest themselves~\cite{Klein:1955zz}. In the Breit frame, the low-energy expansion of the spin-averaged amplitude starts with 
\begin{align}
 \epsilon_f^{\mu \star}T_{\mu\nu}\epsilon_i^\nu&=C_1\,\boldsymbol{\epsilon_f^\star}\cdot\boldsymbol{\epsilon_i}+C_2\,(\boldsymbol{\epsilon_f^\star}\cdot\boldsymbol{n_i}) \,(\boldsymbol{\epsilon_i}\cdot\boldsymbol{n_f})\co\no
C_1 &=-Q_e^2\no
&+\frac{\omega^2}{4\mN^2}(1-z)\{(1+z)(Q_e+\kappa)^2 -Q_e^2\}\no
&+ \omega^2(c_1 + c_2z)+O(\omega^4)\co\no
C_2&=\frac{\omega^2}{4\mN^2}\kappa(2\,Q_e+\kappa)z-\omega^2c_2+O(\omega^4)\co
\label{eq:Compton} 
\end{align}
where ${\boldsymbol n_i}\equiv{\boldsymbol q_i}/\omega, {\boldsymbol n_f}\equiv{\boldsymbol q_f}/\omega$ are the unit vectors in the direction of the initial and final photon momenta, respectively, and $z\equiv{\boldsymbol n_f}\cdot {\boldsymbol n_i}$ (in the Breit frame, the term linear in $\omega$ does not contribute to the spin average). Comparison with the well-known low-energy representation of the Compton scattering amplitude (see for instance~(2.1) and (2.5) in~\cite{Griesshammer}) shows that, up to normalization, the low-energy constants $c_1$ and $c_2$ represent the electric and magnetic polarizabilities, respectively:
\be
\label{eq:Polarizabilities} 
c_1=\frac{\mN}{\alphaem}\alpha_E\co\quad
c_2=\frac{\mN }{\alphaem}\beta_M\fs
\ee
We emphasize that only the sum of the contributions from the Born terms and the polarizabilities manifests itself in Compton scattering. Both the choice of the reference frame and the choice of the Born terms are a matter of  convention~\cite{Scherer:1996ux,Fearing:1996gs,Birse:2012eb}. Accordingly, the literature contains several different variants of the above representation. In the context of the present paper, the relation~\eqref{eq:Polarizabilities} amounts to a definition of the polarizabilities.
\esp

Together with the Born terms, the two polarizabilities $\alpha_E,\beta_M$ determine the low-energy expansion not only of the spin average but of the entire amplitude $\boldsymbol{T}^{\mu\nu}$ up to and including $O(q^2)$. In fact, if not only the photon energies and momenta but the nucleon momenta are also booked as small quantities of $O(q)$, this statement is valid in any reference frame: the leading terms can be written in a manifestly Lorentz invariant manner,
\begin{align}
  \boldsymbol{T}^{\mu\nu}&=\boldsymbol{T}^{\mu\nu}_B+\frac{\mN}{\alphaem}\left\{-\beta_M K^{\mu\nu}_1+(\alpha_E+\beta_M)K^{\mu\nu}_2\right\}\, {\bf 1}\no
  &+O(q^3)\fs
  \label{eq:LEcov}
\end{align}
The quantities $K^{\mu\nu}_1,K^{\mu\nu}_2 $ represent the generalization of the conserved tensors specified in~\eqref{eq:K} to nonforward directions,
\begin{align}
 K_1^{\mu\nu}&= q_i^\mu q_f^\nu-g^{\mu\nu}q_f\cdot q_i\co\no
 K_2^{\mu\nu}&=  \frac{1}{\mN^2}\Big\{(P^\mu q_f^\nu+P^\nu q_i^\mu)P\cdot q\no
 &\qquad-g^{\mu\nu}(P\cdot q)^2-P^\mu P^\nu q_f\cdot q_i\Big\}\co
 \label{eq:Kif}
\end{align}
with  $P=\frac{1}{2}(p_i+p_f)$. Conservation of energy and momentum implies that $P\cdot q_f$ coincides with $P\cdot q_i\equiv P\cdot q$. While the Breit frame formulae~\eqref{eq:LEC}, \eqref{eq:Tbar0} only contain terms of $O(q^2)$, the representation~\eqref{eq:LEcov} of the contributions from the polarizabilities includes higher orders of the low-energy expansion.  

\subsection{Low-energy theorems for $T_1(\nu,q^2)$ and $T_2(\nu,q^2)$}
\label{app:low_energy_theorems}

In the forward direction, the spin average is a Lorentz invariant notion and only two of the four invariant variables are independent: the photon virtualities are the same, $q_i^2 = q_f^2=q^2$, and the momentum transfer $t$ vanishes. The standard variable $\nu=(s-u)/4\mN$ coincides with the frequency in the Breit frame,  $\omega=\nu$.  

The spin average involves the two invariants $T_1,T_2$ defined in~\eqref{eq:normTi} and~\eqref{eq:defVandT}. In the notation of~\eqref{eq:t_n} these amplitudes are given by 
\be
\label{eq:TA} 
T_1=I_2+I_3+I_4+I_5\co
\quad
q^2T_1+\nu^2T_2=I_1\fs
\ee
While for real Compton scattering only the invariants $I_1,I_2$ count, the amplitudes relevant for Compton scattering of virtual photons in the forward direction also pick up a contribution from $I_3,I_4,I_5$.

\bsp
The comparison of the above expressions for the Born terms with the elastic part of the forward amplitudes in~\eqref{eq:Tel} shows that the singularities on the left and right hand sides of~\eqref{eq:TA} are indeed the same, but in the case of $T_1$, the regular parts differ:
\begin{align}
 T_1^\el&= I_2^B+I_3^B+I_4^B+I_5^B+\frac{1}{4\mN^2}F_P^2(q^2)\co\no
T_2^\el&=\frac{1}{\nu^2}\{I_1^B-q^2(I_2^B+I_3^B+I_4^B+I_5^B)\}\fs
\label{eq:T1B}
\end{align}
The difference also shows up when evaluating the relation~\eqref{eq:LEC} in the forward direction, where it implies a low-energy theorem for $T_1$~\cite{Fearing:1996gs,Pachucki:1999zza,Hill:2011wy,Gorchtein:2008aa,Birse:2012eb} as well as one for $T_2$~\cite{Baldin}, 
\begin{align} 
T_1^\inel(0,0)&= -\frac{\kappa^2}{4\mN^2}-\frac{\mN}{\alphaem}\beta_M \co\no
T_2 ^\inel(0,0)&=\frac{\mN}{\alphaem}(\alpha_E+\beta_M)\fs
\end{align}
This demonstrates that, although the amplitudes $T_1(\nu,q^2)$, $T_2(\nu,q^2)$ do not determine the angular distribution of real Compton scattering, they do encode the polarizabilities.
 The calculation described in \app~\ref{app:Derivation} removes the apparent contradiction: the amplitudes $I_3-I_3^B, I_4-I_4^B, I_5$ all disappear if the photon energy is set equal to zero.  
We repeat that the decomposition of the amplitude into a contribution generated by the elastic singularities and a remainder only becomes unique if the asymptotic behaviour is specified. The extra term in $T_1$ arises because the recipe used above to specify the Born terms implies that the amplitudes $I_n^B$ do not tend to zero when $\nu\rightarrow\infty$. In the above analysis of non-forward Compton scattering, the asymptotic behaviour does not play any role -- accordingly, the regular parts of the Born terms used in that analysis are without physical significance. In contrast, the decomposition of the forward amplitudes into an elastic and an inelastic part set up in \sec~\ref{sec:Elastic} does invoke the asymptotic behaviour.  
It implies that the polarizabilities do pick up a contribution from the elastic singularities.  
\esp

\section{Causality}
\label{app:Causality}

\bsp
The structure functions $V_1(\nu,q^2), V_2(\nu,q^2)$ are experimentally accessible only in the space-like region, $q^2\leq 0$. As discussed in  detail in~\cite{GL1975,LO}, causality -- the fact that the current commutator vanishes outside the light-cone -- very strongly constrains their continuation into the time-like region.
General properties of causal functions are described in~\cite{BogoliubovVladimirov,BrosEpsteinGlaser} and
explicit representations that manifestly incorporate causality~\cite{JostLehmann,Dyson,DGSvertex,DGS} are available. These have been used, in particular, in the analysis of the structure functions at high energies~\cite{Brown}. 
The further representation constructed in \app~\ref{app:New} shows that the contributions generated by the elastic intermediate states or individual resonances can be written in manifestly causal form.   
\esp

In the present context, the key statement\footnote{See theorem 2 in~\cite{JostLehmann} and  theorem A in~\cite{Dyson}.}  is that the continuation is uniquely determined up to a polynomial in the variable $\nu$:
\be 
\label{eq:Vfp}
V(\nu,q^2)= \epsilon(\nu)\sum_{n=0}^N \sigma_n(q^2) \nu^{2n}\co
\ee 
where the coefficients $\sigma_n(q^2)$ vanish
for $q^2\leq 0$.
In Regge language, such contributions represent fixed poles in the angular momentum plane, located at integer values of the angular momentum.

Regge asymptotics  excludes fixed poles in $V_2(\nu,q^2)$, but a term with $n=0$, that is a fixed pole with $J=0$,  is not a priori ruled out in $V_1(\nu,q^2)$: 
\be 
\label{eq:V1fixedpole}
V_1^\text{fp}(\nu,q^2)=\epsilon(\nu)\sigma(q^2)\fs
\ee
A term of this form is also consistent with the short distance properties of QCD, which ensure that, in the Bjorken limit, where $\nu$ and $q^2$ both become large, the structure functions tend to zero -- this merely imposes a constraint on the asymptotic behaviour, which in particular requires that $\sigma(s)$ disappears when $s$ becomes large. If a fixed pole were present in $V_1(\nu,q^2)$, it would not show up in the electron cross sections, but would affect
the time-ordered amplitude, through the term
\be
\label{eq:T1fixedpole}
T_1^\text{fp}(\nu,q^2)=\int_0^\infty \frac{ds\, \sigma(s)}{s-q^2-i\epsilon}  \fs
\ee
Accordingly, a formula that expresses the electromagnetic self-energy  in terms of the electron cross sections could then not be given, nor would it be possible to express the polarizabilities of the nucleon in terms of these cross sections.

\bsp
The analysis of~\cite{GL1975} is based on the assumption that the matrix element of the current commutator is free of fixed poles, so that the electron cross sections unambiguously determine the structure functions $V_1(\nu,q^2)$, $V_2(\nu,q^2)$, not only in the space-like region, but also for time-like momenta. The short distance properties of QCD ensure that there is then no ambiguity in $T_1(\nu,q^2)$, $T_2(\nu,q^2)$ either: the electron cross sections fully determine these. Accordingly, the electromagnetic self-energy as well as the polarizabilities of the nucleon  are determined by these cross sections, at least in principle. 
\esp

\section{A new causal representation}
\label{app:New}

Consider the product of two retarded propagators
\begin{align}
t^\ret(p,q)&= \frac{1}{\mu_1^2-(q^0+i \epsilon)^2+{\bf q}^{2}}\no
&\times\frac{1}{\mu_2^2-(p^0+q^0+i \epsilon)^2+({\bf p}+{\bf q})^2} \fs
\label{eq:tret}
\end{align}
The Fourier transform of this amplitude,
\be
\label{eq:tildetret}
\tilde{t}^{\ret}(p,x)=\int d^4q e^{-i x\cdot q}t^\ret(p,q)\co
\ee
is given by the convolution of the two propagators in coordinate space. Since these vanish outside the forward light-cone, the same is also true of the convolution: $\tilde{t}^{\ret}(p,x)$ differs from zero only in the forward light-cone,  $x^2\geq 0,\,x^0\geq 0$.

\bsp
This reflects the properties of the integrand in~\eqref{eq:tildetret}, which contains four poles
\begin{align}
 t^\ret(p,q)&=\frac{1}{(q^0-\omega_1+i\epsilon)(q^0+\omega_1+i\epsilon)}\no
 &\times\frac{1}{(q^0+p^0-\omega_2+i\epsilon)(q^0+p^0+\omega_2+i\epsilon)}\co\no
\omega_1&= \sqrt{\mu_1^2+{\bf q}^{2}}\co\quad\omega_2 =\sqrt{\mu_2^2+({\bf q}+{\bf p})^2}\fs
\label{eq:4poles}
\end{align}
All of these occur in the lower half of the $q^0$-plane. The path of integration can therefore be deformed into a segment from $-\infty$ to $-R$, a semi-circle of radius $R$ and a segment from $R$ to $+ \infty$.  If $x^0< 0$, the factor $e^{-i q^0 x^0}$ suppresses the integrand if
$R$ is taken large, except for the segments where the imaginary part of $q^0$ is not large. But there, $t^\ret(p,q)$ is small, of order $1/R^4$. Since the length of these segments is of order $R$, their contributions also tend to zero if $R$ is taken large. Since the integral is path-independent, it vanishes for $x^0< 0$.
\esp

\bsp
The quantity $\tilde{t}^{\ret}(p,x)$ is Lorentz invariant. The time component of the vector $x$ depends on the frame chosen, but  for any point outside the forward light-cone, there is a frame where the time component is negative. This confirms that $\tilde{t}^{\ret}(p,x)$ vanishes outside the forward light-cone.
\esp

\bsp
The advanced version of the amplitude only differs in the sign of the $i \epsilon$ prescription. Like the advanced propagators in coordinate space, the Fourier transform of $t^{\adv}(p,q)$ is different from zero only in the backward light-cone, $x^2\geq 0$, $x^0\leq 0$. The difference between the two,
\be
\label{eq:v}
v(p, q) = \frac{1}{2 \pi i} (t^\ret(p,q) - t^{\adv}(p,q))\co
\ee                                                                      
is therefore causal: the Fourier transform of $v(p, q)$ vanishes outside the light-cone. Since $t^{\adv}(p,q)$ is the complex conjugate of $t^\ret(p,q)$, the function $v(p,q)$ is real. 
\esp

\bsp
For space-like momenta, $v(p,q)$ picks up a contribution only from the poles at $q^0=-p^0\pm \omega_2$:
\be
\label{eq:v space-like}
v(p, q) = \frac{1}{\mu_1^2-q^2 }\delta((p+q)^2-\mu_2^2)\co\quad q^2\leq 0\co\,\nu\geq 0 \fs
\ee
This demonstrates that any function which for space-like momenta can be represented as
\begin{align}
 V(\nu,q^2)&=\int_0^\infty da\int_0^\infty db\; \frac{\rho(a,b)}{a-q^2}\no
 &\times\{\delta(q^2+2\nu \mN-b)-\delta(q^2-2\nu \mN-b)\}\co
 \label{eq:V}
\end{align}
admits a causal continuation into the time-like region (we have replaced $\mu_1^2$ and $\mu_2^2$ by $a=\mu_1^2$, $b=\mu_2^2-\mN^2$, respectively and imposed the condition $b\geq0$, which ensures that, for space-like momenta, the support of $V(\nu,q^2)$ is contained in the physical region, $Q^2\leq 2\mN|\nu|$). There is only one continuation that is free of fixed poles. The corresponding time-ordered amplitude is given by
\begin{align}
T(\nu,q^2)&= \int_0^\infty da\int_0^\infty db\, \frac{\rho(a,b)}{a-q^2-i\epsilon}\no
&\times \Bigg\{\frac{1}{b-q^2+2\nu \mN-i\epsilon}\no
&\qquad+\ \frac{1}{b-q^2-2\nu \mN-i\epsilon} \Bigg\}\fs
\label{eq:T}
\end{align}
This representation, in particular, yields a causal description of the elastic contributions discussed in \sec~\ref{sec:Elastic}. Since the form factors are analytic functions of $t$, the coefficients $v_1^\el$, $v_2^\el$ introduced in~\eqref{eq:Vel} do admit a representation of the form
\be 
v_i^\el(q^2)=\int_0^\infty da\;\frac{\sigma_i(a)}{a-q^2-i\epsilon}\co
\ee
so that the spectral functions $\rho_i(a,b)=\sigma_i(a)\delta(b)$
indeed generate the elastic contributions. 
\esp

\section{Errata in the literature}
\label{app:Comments}

In the present appendix, we rectify a number of incorrect statements made in the literature~\cite{WalkerLoud:2010qq,WalkerLoud:2012bg,WalkerLoud:2012en,WalkerLoud:2013yua} about the work of Gasser and Leutwyler~\cite{GL1975}. Unfortunately, some of these already propagated~\cite{McGovern:2012ew,Thomas:2014dxa,Erben:2014hza}.

The only deficiency of the analysis reported in~\cite{GL1975} which we are aware of concerns the evaluation of the contributions arising from the deep inelastic region: the violations of Bjorken scaling are not accounted for. In particular, if the ratio $x=Q^2/2\mN\nu$ is kept fixed, the longitudinal structure function $F_{\indL}=F_2-2 x F_1$ is assumed to tend to zero at high energies, in inverse proportion to $Q^2$. In QCD, $F_{\indL}$ only disappears in proportion to $\alpha_s\propto 1/\log(Q^2)$. The calculation yet needs to be improved to account for the scaling violations, but we doubt that this will significantly affect the numerics -- in~\cite{GL1975}, the contribution from the entire deep inelastic region was found to be small. 

The essence of the analysis in~\cite{GL1975} is recapitulated in \sec~\ref{sec:Determination}:  Reggeon dominance implies that the subtraction function is uniquely determined by the cross section of the reaction $e+p\rightarrow e+X$. 
In \cite{WalkerLoud:2010qq}, for instance, this crucial point is not addressed at all. Instead, the paper contains the following statement:\\
{\it In their work, they acknowledged the need for a subtracted dispersion relation but proceeded to ignore this issue, as the subtraction constant could not be computed.}\\
Quite the contrary, the analysis in~\cite{GL1975} not only includes an explicit evaluation of the subtraction occurring in the fixed-$q^2$ dispersion relation for $T_1$, but contains a detailed discussion of the matter in a separate section, entitled ``Regge poles, fixed poles, subtractions and the like.''

In~\cite{WalkerLoud:2012bg} the authors write:\\
{\it In}~\cite{GL1975} {\it it was claimed that the {\it elastic} contributions to $t_1$ could be evaluated with an unsubtracted dispersive analysis. However, performing an unsubtracted dispersive analysis of the elastic contributions [\ldots]  leads to inconsistent results.}\\
This conclusion is obtained by comparing the decomposition of $T^{\mu\nu}(p,q)$ specified in \app~\ref{app:Notation} ($T_1=-\frac{1}{2}t_1$, $T_2=\frac{1}{2}t_2$ in their notation) with an alternative decomposition, introduced by the authors as $\tilde{T}_1=2q^2 T_1+2\nu^2 T_2$, $\tilde{T}_2 = -2q^2 T_2$. While the amplitudes $T_1,T_2$ are free of kinematic singularities and zeros, $\tilde T_1,\tilde T_2$ are not. That is why the authors run into an inconsistency:  while $T_2$ does obey an unsubtracted dispersion relation, the dispersive representation for $\nu^2 T_2$ requires a subtraction. The need for a subtraction also shows up in the asymptotic behaviour of the elastic contributions in~\eqref{eq:Tel}, which implies that $\tilde{T}_1^\el$ does not disappear when $\nu\rightarrow  \infty$ and can therefore not possibly obey an unsubtracted dispersion relation. The calculation described by the authors merely shows that amplitudes with kinematic zeros may fail to obey unsubtracted dispersion relations. It does not demonstrate that the representation of the elastic contributions $T_1^\el,T_2^\el$ in~\cite{GL1975} is incorrect -- in fact, as shown in \app~\ref{app:New}, even taken by itself, that representation for the elastic contributions to $T_1(\nu,q^2)$, $T_2(\nu,q^2)$ obeys all of the constraints imposed by causality.

\bsp
Further, in the conclusion of~\cite{WalkerLoud:2012bg}, the authors state:\\
 {\it A technical oversight in the evaluation of the elastic contribution was highlighted resulting in a larger central value than previously obtained} \cite{GL1975}.\\
The claim of a 'technical oversight' suggests that the value $0.76\MeV$ for the elastic contribution obtained in~\cite{GL1975} is incorrect and needs to be replaced by $1.39(2)\MeV$.  That, however, is not the case. The number quoted in~\cite{GL1975} concerns the full contribution of the elastic intermediate states to the Cottingham formula, which includes the elastic contribution to the subtraction function. In~\cite{WalkerLoud:2012bg}, the elastic part of the mass-shift is instead identified with the contribution from the elastic states to the {\it subtracted} dispersion integral. In fact, the authors notice that their model for the subtraction function also contains elastic contributions and that if these are accounted for, the claimed discrepancy in the value of the elastic contribution to the self-energy difference disappears.\footnote{Incidentally, although their amended expression for the elastic part of the self-energy is numerically close to the contribution generated by the elastic singularities, this is only approximately so: while these singularities are proportional to $G_E^2$ and $G_M^2$, their expression for the elastic part of the mass-shift in addition involves an integral over $G_E G_M$.} This demonstrates that their claim is wrong and that the evaluation in~\cite{GL1975} is correct.
\esp

\bsp
The incorrect  statements concerning the analysis in~\cite{GL1975} are iterated elsewhere. For instance, the following claims are made:\\ 
\cite{WalkerLoud:2012en}: {\it One is lead to conclude that the unknown subtraction function can not be
evaded and the evaluation of $\delta \mg^\gamma$ in Ref.} \cite{GL1975} {\it is not correct.}\\
\cite{WalkerLoud:2013yua}:
{\it This work uncovered a technical oversight in work of Gasser and Leutwyler} \cite{GL1975,GL1982} {\it related to a subtracted dispersion integral, which unfortunately invalidates their result.}
\esp

Finally we point to an erratum in~\cite{WalkerLoud:2012en}:\\\
{\it In Ref.}~\cite{GL1975} {\it an argument to evade the subtraction function based on the parton model was presented. However, as was  first noted in Ref.}\cite{Collins}, {\it the argument was based on false
assumptions about the scaling violations of the Callan-Gross relation} \cite{CallanGross}.\\
The necessity of a subtraction arises from the Regge behaviour at fixed $q^2$. The analysis in~\cite{GL1975} relies on standard Regge behaviour and hence necessarily involves a subtracted dispersion relation for $T_1$. Scaling violations concern the behaviour of the structure functions in the deep inelastic region, where $q^2$ becomes large. It is true that in~\cite{GL1975}, the scaling violations are ignored, but to claim that this is used in~\cite{GL1975} as {\it an argument to evade the subtraction function} is plain wrong: a subtraction was made and the subtraction function was determined.

\end{document}